\documentclass[paper]{JHEP} 
\input{epsf.sty}
\usepackage{epsfig}

\newcommand{\NPB}[3]{\emph{ Nucl.~Phys.} \textbf{B#1} (#2) #3}   
\newcommand{\PLB}[3]{\emph{ Phys.~Lett.} \textbf{B#1} (#2) #3}   
\newcommand{\PRD}[3]{\emph{ Phys.~Rev.} \textbf{D#1} (#2) #3}   
\newcommand{\PRL}[3]{\emph{ Phys.~Rev.~Lett.} \textbf{#1} (#2) #3}

\newcommand{\JHEP}[3]{\emph{JHEP} \textbf{#1} (#2) #3}
\def\dalemb#1#2{{\vbox{\hrule height .#2pt
        \hbox{\vrule width.#2pt height#1pt \kern#1pt
                \vrule width.#2pt}
        \hrule height.#2pt}}}

\let\a=\alpha    
    
    \let\p=\pi 
     
      \let\G=\Gamma   \let\L=\Lambda
    \let\F=\Phi

 \def\bd{\begin{document}} \def\ed{\end{document}}
\def\ds{\documentstyle} \let\fr=\frac \let\bl=\bigl \let\br=\bigr
\let\Br=\Bigr \let\Bl=\Bigl 
\let\bm=\bibitem
\let\na=\nabla
\let\pa=\partial \let\ov=\overline
\def\ie{{\it i.e.\ }} 
\def\tr{{\mbox{\rm tr}}}
\newcommand{\be}{\begin{equation}}
\newcommand{\ee}{\end{equation}}
\newcommand{\beba}{\begin{equation}\begin{array}{lcl}}
\newcommand{\eaee}{\end{array}\end{equation}}
\newcommand{\bea}{\begin{eqnarray}}
\newcommand{\eea}{\end{eqnarray}}
\newcommand{\ba}{\begin{array}}
\newcommand{\ea}{\end{array}}
\newcommand{\td}{\tilde}
\newcommand{\norsl}{\normalsize\sl}
\newcommand{\ns}{\normalsize}
\newcommand{\refs}[1]{(\ref{#1})}
\def\simlt{\mathrel{\lower2.5pt\vbox{\lineskip=0pt\baselineskip=0pt
           \hbox{$<$}\hbox{$\sim$}}}}
\def\simgt{\mathrel{\lower2.5pt\vbox{\lineskip=0pt\baselineskip=0pt
           \hbox{$>$}\hbox{$\sim$}}}}
\def\A{{\cal A}}
\def\a{{\mathcal a}}
\def\V{{\cal V}}
\def\F{{\cal F}}
\def\L{{\cal L}}
\def\p{{\mathcal \phi}}

\title{Contact interactions in D-brane models}

\author{I. Antoniadis\thanks{On
leave from {\it Centre de Physique Th{\'e}orique, Ecole Polytechnique,
F-91128 Palaiseau} }
\\CERN Theory Division
 CH-1211, Gen{\`e}ve 23, Switzerland }  

\author{ K. Benakli \\
CERN Theory Division
 CH-1211, Gen{\`e}ve 23, Switzerland \\$~~~~~~~~~~~~~~~~~~~~~~~~~~~~~~$ and\\ 
Department of Physics, McGill University, Montr\'eal, QC, H3A 2TS, Canada}

\author{ A. Laugier\\
Centre de Physique Th{\'e}orique, Ecole Polytechnique,
91128 Palaiseau, France\\
(Unit\'e Mixte du CNRS et de l'EP, UMR 7644)}

\abstract{We compute the tree-level four-point scattering amplitudes in
string models where matter fields live on D-brane intersections. 
Extracting the contribution of massless modes, we are left with dimension-six four-fermion operators which in general receive contributions from three
different sources: exchange of massive Kaluza--Klein excitations, winding
modes and string oscillator states. We compute their coefficients
and extract new bounds on the string scale in the brane-world scenario.
This is contrasted with the situation where matter fields arise from open
strings with both ends confined on the same collection of D-branes, in which
case the exchange of massive string modes leads to dimension-eight operators
that have been studied in the past. When matter fields live on brane
intersections, the presence of dimension-six operators increases the lower
bound on the string scale to 2--3 TeV, independently of the number of
large extra dimensions.}

\preprint{hep-th/0011281\\CERN-TH/2000-344\\CPHT-S092-11-00\\McGill-00-32
}
\keywords{strings, D-branes,  extra dimensions}
\begin{document}

\section{Introduction}

The world-volume of a D$p$-brane \cite{revue1} 
can be described as a $p+1$ dimensional
space-time where open strings can end and propagate. This free 
propagation of the string endpoints is described by imposing Neumann (N) 
boundary conditions for the world-sheet fields. In the remaining $9-p$ transverse
dimensions the endpoints of these open strings are confined at the
location of the brane and satisfy Dirichlet (D) conditions. The existence of a
string world-sheet description for the D-brane dynamics of world-volume
and bulk states provides a powerful tool that allows to compute
 higher derivative corrections as well as  string loop corrections.

A generic configuration of D-branes contains branes
with world-volumes of different dimensionalities.
Along a given direction, open strings stretched between these 
different branes could satisfy NN, DD, or ND boundary conditions, depending on
whether this direction is part of the world-volume of both 
D-branes, transverse to both of them, or along the world-volume of the one but 
transverse to the other, respectively.
   
Orbifold compactifications of type I strings \cite{orientifold,models}
provide the simplest  framework allowing to investigate  the
possibility that standard model fieldsare identified  with degrees of
freedom confined on D-branes. Search for such realistic models
indicates that the configuration of branes should contain at least two
sets, for instance D3 and D7 branes (or a T-dual configuration).
While gauge particles are associated to NN or DD open strings with
both ends on the same set of branes, chiral matter fields are usually
localized on brane intersections and satisfy ND boundary conditions in
some internal directions. Besides providing chirality, this choice
avoids possible phenomenological problems related to the fact that if
all matter fields were associated to NN or DD strings as the gauge
fields, the lowest Kaluza--Klein (KK) excitations would be stable. This
would lead, in particular, to stable charged states with TeV masses in
models with low compactification or string scales~\cite{AB,AAB}.

Confining standard model states on D-branes 
opens new  possibilities for phenomenological applications
of string theory as it becomes possible to lower
its fundamental scale much below the Planck mass \cite{Ant,ADD,more}. The
most spectacular of these proposals are scenaria with the string scale
$M_s \equiv l_s^{-1}$ lying at energies as low as a few TeV, where a
plethora of new phenomena could be observable at future colliders.
In order to derive constraints on these models, it is important to study
deviations from the standard model expectations for low-energy
cross-sections.

Such deviations manifest themselves for instance as
the appearance of higher dimensional operators. Here we discuss these
operators in the case of orbifolds with two types of branes that we take for
concreteness as D3 and D7 branes. We then find that in addition to the
dimension-six four-fermion operators induced by the exchange of KK states or winding
modes, there are also dimension-six operators induced by massive string
oscillators. All such contact interactions appear when standard
model matter fields are identified with massless modes of open strings stretched
between the D3 and D7 branes (37 strings), satisfying ND boundary conditions along
four internal directions. These effects dominate the dimension-eight operators 
\cite{AAB,Peskin} obtained for the case where matter fields are identified with
massless modes of DD strings (33 or 77 strings).

In the context of the above framework, one can assign two different embeddings of
the Standard Model:
\begin{enumerate}
\item One possibility is that all observable gauge fields live on the world-volume of
D3 branes, while the D7 branes are in general extended (partially) in the bulk. 
If matter fields come from open strings with both ends on the D3 branes, the presence
of D7 branes is irrelevant for our purposes and this case is reduced to the one
studied in ref. \cite{Peskin} and summarized in section 3.1. The only possible source
of massive exchanges are string oscillator modes that lead in this case to dimension-eight effective operators.
\item Other possibilities are obtained when the Standard Model is splitted in two factors living correspondingly on the D3 and D7 branes which are transverse
simultaneously to the two remaining extra dimensions of the bulk (of mm
size) \cite{AAB,kiritsis}. For instance, strong and weak interactions could be confined on the
D3 and D7 branes, respectively, and the size of the four-dimensional internal
volume along the D7 branes could account for the weakness of the $SU(2)$ relative to
the $SU(3)$ couplings.
\end{enumerate}

As we mentioned above, in both cases, the presence of matter fields
originating from 37 strings leads to dimension-six operators which
receive contributions from exchanges of KK modes of the 77 strings
(having both ends on the D7 branes). In scenario (1), they are
associated to new (exotic) degrees of freedom with new interactions to
Standard Model states. Since these are model dependent and presumably
suppressed in realistic models, they can not be used to obtain model
independent experimental bounds. On the other hand, in scenario (2),
the KK states correspond to heavy excitations of Standard Model gauge
degrees of freedom. The effects of their exchanges should then be used
to obtain experimental bounds that have been studied in the past
\cite{AB,KK1,KK2}.

In addition to the KK excitations of 77 states, there are string
winding modes around the D7 brane world-volume compact directions
which carry D3 brane quantum numbers and can be exchanged among 37
states. However, their contribution is exponentially suppressed in the
large volume limit and can be neglected. The remaining contributions
to dimension-six operators come from the exchange of massive
oscillators of 33 and 77 strings and can be extracted explicitly.
They can be used to derive  model independent bounds on
the string scale, in the context of our framework.

It is important to note that our results remain valid in
non-supersymmetric models with brane supersymmetry breaking, where
some D-branes are replaced by anti-D-branes of the same type
\cite{antibrane}.  Supersymmetry is then broken (to lowest order) on
the world-volume of the anti-D-branes.

Our paper is organized as follows. In section 2, we recall the basic
properties of D-brane models and define our framework. In section 3,
we compute all possible four-fermion tree-level scattering amplitudes,
while in section 4, we extract the four-fermion contact
interactions. For completeness, we also present the computation of the
four-scalar amplitude in the Appendix. The reader who is not
interested in the detailed derivations can skip sections 3 and 4 and
go directly to section 5, which summarizes our results. Finally, in
section 6, we perform the numerical analysis and derive the bounds on
the string scale.

\section {The D-brane model}

In this work we consider orbifold compactifications of type I string
theory to four dimensions.  Gauge degrees of freedom are described by
open strings with ends confined on sets of D-branes.  A Dp-brane is
obtained by imposing Dirichlet  (D) boundary conditions for the
endpoints  of open strings along the $9-p$ directions transverse to
the brane and  Neumann (N) conditions along the longitudinal  $p+1$
(space + time) directions. Generically, $N$ coincident D$p$-branes
give rise to a $U(N)$ gauge group, that can be reduced to the
orthogonal or symplectic maximal subgroup by the orientifold
projection.  

In the supersymmetric case, such a generic vacuum contains
four types of Dp-branes which upon appropriate T-dualities can be
mapped to one set of D3 branes and three different sets of D7 branes:
\begin{table}[h]
\begin{tabular}{cccccccccccc}
        &0&1&2&3&4&5&6&7&8&9&located at \\
 D$3_I$ branes:&N&N&N&N&D&D&D&D&D&D&$X_{i+3} =a^{I}_{i} R_i$\\
 D$7_1$ branes:&N&N&N&N&D&D&N&N&N&N&$X_4=X_5=0$ \\
 D$7_2$ branes:&N&N&N&N&N&N&D&D&N&N&$X_6=X_7=0$ \\
 D$7_3$ branes:&N&N&N&N&N&N&N&N&D&D&$X_8=X_9=0$ \\ 
\end{tabular}
\begin{center}
{\bf Table 1}
\end{center}
\label{defbranes}
\end{table}

\vspace{-0.7cm}
\noindent
where for simplicity, the six internal coordinates  $X_{i+3}$,
$i=1,\cdots,6$  are compactified on a  six-dimensional torus with
radii $R_{i}$. For generality, we have allowed the D3 branes to be
localized at different points in the transverse directions
parameterized by $a^{I}_{i}R_i$ where $I$ labels the 
branes. Their separations break the gauge group accordingly.
Recently,  non-supersymmetric models with supersymmetry broken on the
branes  have been constructed \cite{antibrane}. These models contain 
the same type of
branes with  the difference that some of them are replaced by
anti-D-branes  (having opposite Ramond-Ramond charges). Their numbers
are constrained by the  cancellation of all Ramond-Ramond
charges in the compact internal space, as required by the Gauss law.

The open string spectra contain two kinds of states: gauge fields
arise from  strings with both ends on the same set of branes, which
therefore satisfy the same boundary condition for both ends (NN or
DD). Matter fields, on the other hand, can also arise from strings
stretched between two different sets of branes. These strings have ND
boundary conditions along four of the internal directions. They
transform in the bifundamental representation of the two gauge groups
associated to the two sets of branes. Moreover, orientation reversal  of
these strings amounts to complex conjugation. The mass  formulae for
the various cases are:

\begin{itemize}

\item  Open strings with one end on the  D$3_I$ and the other on the 
D$3_J$ branes: 
\be
M^2_{33}=
\sum_{i=1}^{6}\frac {(m_i + a^{IJ}_i)^2R_{ i}^2}{l_s^4}+ 
\frac {N} {l_s^2}\, ,
\label{mopen1}
\ee
where $a^{IJ}_i =  a^{J}_i - a^{I}_i$ parameterizes the separation
$a^{IJ}_i R_{ i}$ of the branes along the direction $i$ and $N$ is
the integer string oscillator number. These states have no KK excitations 
but have winding modes.

\item Open strings with both ends on the same set of  D7 branes:

\be
M^2_{77}= \sum_{ \perp}\frac {n_\perp^2R_{ \perp}^2}{l_s^4} +
\sum_{ \parallel} \frac {n_\parallel^2}{R_{\parallel }^2}+ 
\frac {N} {l_s^2}\, .
\label{mopen2}
\ee
These states have KK excitations along the four longitudinal ($\parallel$) internal dimensions 
 and winding modes along the remaining two transverse ($\perp$) ones.

\item Open strings with one end on the  D3 and the other on a 
D7 set of branes:
 
\be
M^2_{3 7}= \sum_{ \perp}\frac {n_\perp^2R_{\perp }^2}{l_s^4}+ 
\frac {N} {2 l_s^2}\, ,
\label{mopen3}
\ee
These states have winding modes only along the two directions transverse to 
both sets
of branes. The simultaneous absence of  KK and winding modes along the 
remaining four internal directions  associated to ND boundary conditions 
reflects the property that the endpoints of these
strings live in the intersection of D3 and D7 branes.

\item Open strings stretched between two different sets of D7 branes:
\be
M^2_{77'}= \sum_{ \parallel } \frac {n_\parallel^2}{R_\parallel^2} +
\frac {N} {2 l_s^2 }\, .
\label{mopen4}
\ee
These states live in the intersection of  D7 and D7' 
and have only KK excitations along the two directions longitudinal to both
sets of branes.

\end{itemize}

Note that the states 77 and 77' are related by T-duality to those of
33 and 37 states, respectively. The open string endpoints  carry also
gauge indices  that  are described by Chan-Paton matrices  $\lambda$.

Besides the open string spectrum, there are closed strings
 propagating in the whole ten-dimensional space.
Their massless modes describe particles with gravitational coupling 
to matter which include the graviton, dilaton  as well as some model 
dependent states, such as graviphotons and moduli fields. Their masses are 
given by:
\be
M^2_{closed}= \sum_{i=1}^{6} \frac {m_i^2}{R_{ i}^2} 
+\sum_{i=1}^{6} \frac {n_i^2R_{ i}^2}{l_s^4} +
\frac {4 N} {l_s^2}\, .
\label{mclosed1}
\ee

\section{ Four-fermion scattering amplitudes}

In this section, we compute tree-level four-point amplitudes involving 
massless fer\-mions. Each external state is described by
 the insertion of a vertex operator $\V^{(a)}$
$a=1,\cdots,4$  on the boundary of the world-sheet surface with  topology 
of a disk. Using  conformal transformations, the disk can be mapped to the
upper-half complex plane and the world-sheet  boundary 
 to the real line.  There are many possible orderings of the  4
vertex operators along the line at the positions $x_i$.  Using
$SL(2,R)$ conformal invariance one can fix these positions  at
$0,x,1,\infty$. The corresponding four-point  ordered
amplitude is:  
\bea 
\! \!(\!2\pi\!)^4 \! \delta^{(4)}(\sum_a k_a)\! A(1,2,3,4)\! =\! \frac{-i}{g_s l_s^4}\!
\int_{0}^{1}\!\! dx\! \left<\! {\cal V}^{(1)} (0,\!k_1) {\cal
V}^{(2)} (x,\!k_2) {\cal V}^{(3)} (1,\!k_3) {\cal V}^{(4)} (\infty,\!k_4)\!\right>\
\eea 
where $k_i$ are the space-time momenta.   The total scattering  
amplitude $\A_{total}(1,2,3,4)$ is
obtained by summing over  all  possible orderings of the vertices.
Defining $\A(\!1,2,3,4\!)$ $=$ $A(1,2,3,4) + A(4,3,2,1)$, one has:
\be
\A_{total}(1,2,3,4) = \A (1,2,3,4) + \A (1,3,2,4) + \A (1,2,4,3)
\label{Atotal}
\ee
These amplitudes depend on kinematical invariants that can be expressed in terms of
the  Mandelstam variables:
\be
s= -(k_1+k_2)^2 \, , \, \, \, \, \, t= -(k_2+k_3)^2 \, , \, \, \, \, \, 
u=-(k_1+k_3)^2
\label{Mand}
\ee

\subsection {Four fermions from  DD open strings}

The simplest and well known case \cite{myers,Peskin} is the  interaction of four 
massless fermions
living on  the same set of D3  or D7 branes. The corresponding
vertex operator associated with  an open string 
stretched between the  coincident branes $I$ and $J$  is 
given by (in the $-1/2$ ghost picture):
\be
{\cal V}^{(a)}_{DD}(x_a,k_a) = \sqrt {2 g_s\, l_s^3}\,  \, \lambda^a_{IJ}
 u^{(a)}_\alpha S^\alpha_{(10)} \, e^{-\p/2} e^{i k_a\cdot X} (x_a)\, ,
\ee
with $k_a^2 =0$, $\lambda$ the corresponding Chan-Paton matrix and $u^{(a)}_\alpha$ the spinor polarizations.  
Here $\phi$ is the superconformal ghost field and 
$S^\alpha_{(10)}$ the spin fields of $SO(10)$ \cite{CFT}:

\be
S^\alpha_{(10)} = :e^{ \frac{i}{2}{\vec {\tilde t}_\alpha}\cdot{\vec  H}}:, \qquad 
{\vec H} =\{ H_1,\cdots,H_5 \}, \qquad {\vec {\tilde t}_\alpha} = 
\{ \pm,\pm,\pm, \pm ,\pm \},
\ee
where $H_i$ are two-dimensional bosonic fields and there is no correlation 
among the $\pm$ in the different positions. The GSO
 projection  projects half of the components of the spinors 
$S^\alpha_{(10)}$, keeping for 
instance those with $\prod_{i=1}^5 {\tilde t_i} = + $.  The simultaneous presence of 
D3 and D7 branes results in further projecting out half of the spinors 
by an appropriate implementation of Dirichlet boundary conditions (for instance
$t_2t_3=-$).

In order to introduce also a flavor (family) index let us consider a 
concrete example of compactification to four dimensions based on the $Z_3$
orientifold which allows to construct three-family models \cite{models}.
 The $Z_3$ orbifold group action transforms the three complex 
coordinates  $z_k=X_{k+3} +i X_{k+4}$  and  the world-sheet fields $H_i$
as:
\bea
z_k &&\longrightarrow  g\, z_k =
e^{\frac {i 2 \pi}{3}} z_k\nonumber \\
\{ H_1, H_2,H_3, H_4, H_5 \} && \to 
\{ H_1,H_2,H_3 +{\frac { 2 \pi}{3}},H_4+ {\frac{ 2 \pi}{3}},H_5 + {\frac{ 2 \pi}{3}}\}
\eea
The vector   ${\vec {\tilde t}_{\alpha}}$  can be decomposed as
  ${\vec {\tilde t}_{\alpha}} = \{ 
{\vec s_{\alpha}}, {\vec t_{\alpha}} \}$ with 
${\vec t_{\alpha}} =\{t_1,t_2, t_3\}$ while
 ${\vec s_{\alpha}} =\{s_1,s_2\}$ defines the  four-dimensional helicity. 
For instance
left-handed two-component spinors correspond to ${\vec s_{\alpha}} =
\{+,+\}$ and  $\{-,-\}$, while right-handed ones 
are associated with ${\vec s^{\dot\alpha}}=\{+,-\}$ and  $\{-,+\}$.
 Only one left-handed two-component spin field  
corresponding to  ${\vec t} = \{+,+,+\}$ and ${\vec s_{\alpha}} =
\{+,+\}$, $\{-,-\}$ 
is left invariant by the $Z_3$ action. It gives rise to the 
gauginos in four-dimensional $N=1$ supersymmetric theories.
On the other hand, through the tensor product of the remaining spin-fields 
with Chan-Paton matrices that transform with  opposite phases, one obtains
three 
families of matter fermions associated with ${\vec t} = \{+,-,-\}$, 
$ \{-,+,-\}$ and $\{-,-,+\}$.

A straightforward computation of the ordered amplitude gives:
\bea
\A(1,2,3,4)& =& -2 g_s \, l_s^2 \,  \tr[ \lambda^1 \lambda^2 \lambda^3 \lambda^4 + 
                  \lambda^4 \lambda^3 \lambda^2 \lambda^1 ] \, \,
\int_0^1 dx \, \,  x^{-1 -s\, l_s^2} \,  \, (1-x)^{-1 -t\, l_s^2} \, \,  \nonumber \\
&\times& [   {\bar u}^{(1)} \gamma^{(10)}_{M} u^{(2)} 
{\bar u}^{(4)} \gamma^{(10)M} u^{(3)}   (1-x)  +
 {\bar u}^{(1)} \gamma^{(10)}_{M} u^{(4)}  
{\bar u}^{(2)} \gamma^{(10)M} u^{(3)} x  ]\, , \nonumber\\
\label{known}
\eea where $\gamma^{(10)}_{M}$ are the $\gamma$-matrices in 10
dimensions. As an example we will consider all four fermions of the
same flavor, more precisely two with the same internal helicity ${\vec t}$
and two with the opposite. In this
case $\gamma^{(10)}_{M}$ are reduced to  the usual four-dimensional
$\gamma$-matrices $\gamma^{\mu}$. The corresponding  ordered
amplitudes for various four-dimensional helicities are then  given by
\cite{Peskin}: 
\bea  
\A(1_L^-, 2_R^+ , 3_L^-, 4_R^+) &=& - 4 \,  g_s\,
\tr[ \lambda^1 \lambda^2 \lambda^3 \lambda^4 +  \lambda^4 \lambda^3
\lambda^2 \lambda^1 ] \cdot \frac {u^2} {st} \F(s,t) \ , \\  \A(1_L^-,
2_R^+ , 3_R^-, 4_L^+) &=&  - 4 \, g_s \,   \tr[ \lambda^1 \lambda^2
\lambda^3 \lambda^4 +  \lambda^4 \lambda^3 \lambda^2 \lambda^1 ] \cdot
{t\over s} \F(s,t) \ , \\  \A(1_L^-, 2_L^+ , 3_L^- ,4_L^+) &=&  - 4 \,
g_s\,  \tr[ \lambda^1 \lambda^2 \lambda^3 \lambda^4 +  \lambda^4
\lambda^3 \lambda^2 \lambda^1 ] \cdot {s\over t} \F(s,t) \ ,  \eea
while the other  non-vanishing  processes can be trivially obtained by
parity reflection. All  the amplitudes above  are expressed in terms
of the effective field theory result multiplied by the string
form-factor  \cite{Venez} \be \F(s,t) = {\G(1-l_s^2 s) \G(1-l_s^2 t)
\over \G(1-l_s^2s-l_s^2t)} .  \ee In the low energy limit $|s \,
l_s^2| \ll 1$, $|t\,  l_s^2| \ll 1$, it can be expanded as:  \be
\F(s,t) = 1 - {\pi^2\over 6}  {st\over M_s^4} + \cdots
\label{dim81}
\ee 
Note that the first correction to the  field theory result,
obtained from the  term proportional to $st$ in eq. (\ref{dim81}),
corresponds to dimension-eight  effective operators.

The above result remains valid in non-supersymmetric models where the
fermions  arise from open strings ending on the same set of
anti-branes.

\subsection {Four ND fermions from  two sets of branes}

The  vertex operator describing the emission of a  massless fermion
originating  from an open string  stretched between  the D$7$-brane
$j$ and  the D3 brane $I$ is  given by: 
\be {\cal V}^{(a)}_{ND}
(x_a,k_a) =  2^{\frac{1}{4}}\sqrt {g_s\, l_s^3}\,  \,  \lambda_{j_a
I_a} u^{(a)}_\alpha S^\alpha_{(6)} \, \prod_{i=1}^4 \sigma^i_{\pm} \,
e^{-\phi/2}  e^{i   k_a\cdot X} (x_a) .  \ee 
where $\sigma^i_{\pm}$ is
the $Z_2$-twist operator acting on the direction  $i$, with conformal
dimension $1/16$; $\sigma_{+}$  changes  a Dirichlet  to Neumann
boundary condition transforming a D$p$ to a D$(p+1)$-brane while the
anti-twist $\sigma_{-}$  does the reverse, transforming a D$(p+1)$ to
a D$p$-brane. Each  ND vertex  operator contains four such twist
fields. $S^\alpha_{(6)}$ are the spin fields of $SO(6)$: 
\be
S^\alpha_{(6)} = :e^{ \frac{i}{2}{\vec {t}_\alpha}{\vec H}}:\,
, \qquad  {\vec H} =\{ H_1,H_2,H_3 \}, \qquad {\vec {t}_\alpha}
=  \{ \pm,\pm,\pm \}.  
\ee 
The GSO projection  projects half of the
components of the spinors  $S^\alpha_{(6)}$ keeping for  instance
those with $\prod_{i=1}^3 {t}_i = + $.   Here  ${\vec {t}_{\alpha}}\equiv \{ 
{\vec s_{\alpha}}, s_1 s_2 \}$ with ${\vec s_{\alpha}} =\{s_1,s_2\}$ giving the
four-dimensional helicity as for the DD case. The Chan-Paton matrices $\lambda$
transform in the bifundamental representation of the D3 and D7 gauge groups, in
the simplest case $U(N_3)\times U(N_7)$ respectively, and can be represented by
$(N_3+N_7)\times (N_3+N_7)$ matrices with one non-vanishing off-diagonal element
in a complex basis.

Let us first consider the amplitude involving four ND fermions from
open strings  stretched between two different sets of branes which we
choose to be D3 and  D7 of the third type in table 1,
transverse to the 8th and 9th directions.  A non-vanishing result
requires at most two sets of coincident D3 branes $I$  and $J$.  The
non-trivial part of the computation involves the correlation function
of two pairs of twist--anti-twist  operators \cite{twist,AB,narain}:
\bea 
\left< \sigma_{+} (0) {\sigma}_{-} (x) {\sigma}_{+} (1)
{\sigma}_{-} (\infty)\right>  =  \frac {[ x(1-x) ]^{-1/8}}{ [F
(x)]^{1/2} }    \sum_{n_i \in {\bf Z}} e^{-  {\pi \tau}  \sum_{i=1}^4
(n_i +a_i^{IJ})^2\,   R_i^2 \, l_s^{-2} },
\label {fourtw}
\eea
where $F(x)=F(1/2,1/2;1;x)$ is the hypergeometric function
\be
F(x) \equiv \int_0^1 dy\,  \, y^{-1/2}\,  \,  (1-y^{-1/2}) \, \, (1-xy)^{-1/2}, \qquad
\tau (x)\equiv  \frac{F(1-x)}{F(x)}, \qquad x\equiv \frac{\theta_2^4}{\theta_3^4}(i \tau), 
\label{xtotau}
\ee
with $\theta$'s the Jacobi theta-functions.

Putting together all correlation functions, the 
ordered four-point amplitude  is given by :
\bea
 \A (1_{j_1 I_1},2_{j_2 I_2},3_{j_3 I_3},4_{j_4 I_4})&=& -   
 { g_s} l_s^2 \int_0^1 dx \, \, x^{-1 -s\, l_s^2}\, \, \,  
(1-x)^{-1 -t\, l_s^2} \, \, \,  \frac {1}{ [F (x)]^2 } \nonumber \\ 
 & \times&  \left[   {\bar u}^{(1)} \gamma_{\mu} u^{(2)} 
{\bar u}^{(4)} \gamma^{\mu} u^{(3)} (1-x) + {\bar u}^{(1)} \gamma_{\mu} 
u^{(4)} {\bar u}^{(2)} \gamma^{\mu} u^{(3)}  x \right ]  \nonumber \\ 
& \times& \{\frac{ \delta_{I_1,{\bar I_2}} \delta_{I_3,{\bar I_4}}
\delta_{{\bar j_1}, j_4} \delta_{j_2,{\bar j_3}}   }
{l_s^{-4} \, \prod_{i=1}^{4} R_i\, }
\sum_{m_i\in {\bf Z}} {e^{2 i \pi \sum_{i=1}^{4} m_i a_i^{I_1I_3}} \, \, e^{ - {\pi} {\tau}\,
\sum_i \frac {  m_i ^2 \, l_s^2 }{ R_i^2}   }}
\nonumber \\ 
&& +  \delta_{j_1,{\bar j_2}} 
\delta_{j_3,{\bar j_4}}
\delta_{{\bar I_1}, I_4} \delta_{I_2,{\bar I_3}} 
\sum_{n_i\in {\bf Z}}  e^{-  {\pi \tau}  \sum_{i=1}^{4} (n_i +a_i^{I_1I_3})^2\,  R_i^2 \, l_s^{-2} } \} 
\label{mastereq}
\eea 
where we made explicit the dependence on Chan-Paton indices. The total
amplitude for the scattering of four  fermions is obtained by summing over
different orderings.  Note that the
$\gamma$-matrices appearing in the second line of (\ref{mastereq}) should be in
general the six-dimensional ones. However, as  in the previous case of DD
fermions we have chosen the internal fermion polarizations such that
 only the four-dimensional matrices appear.

 The
field theory  result is obtained by taking the limit of coincident
vertices  $x \rightarrow 0$ or $x \rightarrow 1$. The behavior of
the special function $F(x)$ in these limits is given by:
\bea
x &\rightarrow& 0 \, : \qquad F(x) \sim 1  \, , 
\qquad \qquad \qquad F(1-x) \sim \frac {1}{\pi} \ln {\frac {\delta}{x}}, \qquad
\tau \to \infty \nonumber \\
x &\rightarrow& 1 \, , \qquad F(x) \sim  \frac {1}{\pi} 
\ln {\frac {\delta}{1- x}} \, : \qquad F(1-x) \sim 1 \, , \qquad \qquad
\tau \to 0,
\label{limtau}
\eea
where $\delta = 2^4$.

The limit $x \rightarrow 0$  leads to: 
\bea
{\lim _{x\rightarrow 0}}\A(1_{j_1 I_1},2_{j_2 I_2},3_{j_3 I_3},4_{j_4 I_4})
 &=&   
  {g_s} \, \, \left[   {\bar u}^{(1)} \gamma_{\mu} u^{(2)} 
{\bar u}^{(4)} \gamma^{\mu} u^{(3)} \right ] \, \, \nonumber \\ 
 \times \left\{  \right. \frac {  \delta_{I_1,{\bar I_2}} \delta_{I_3,{\bar I_4}}
\delta_{{\bar j_1}, j_4} \delta_{j_2,{\bar j_3}}   } 
{ l_s^{-4}\, \prod_{i=1}^{4} R_i}
&& \left[ \frac {1}{s} +\sum_{m_i \in {\bf Z}-\{ 0\}} 
\frac {e^{i  2 \pi \sum_{i=1}^{4} m_i a_i^{I_1I_3}} \, \, 
\delta^{  - \sum_{i=1}^{4} \frac{m_i^2 l_s^2}{ R_i^2}}}
{s - \sum_{i=1}^{4} \frac{m_i^2}{ R_i^2}} \right] \nonumber  \\ + 
\delta_{j_1,{\bar j_2}} 
\delta_{j_3,{\bar j_4}}
\delta_{{\bar I_1}, I_4} \delta_{I_2,{\bar I_3}}  
&& \left[\sum_{n_i\in {\bf Z}} \frac 
{\delta^{-   \sum_{i=1}^{4} (n_i +a_i^{I_1I_3})^2\,  R_i^2 \, l_s^{-2} }}
{s  -   \sum_{i=1}^{4} (n_i +a_i^{I_1I_3})^2\,  R_i^2 \, l_s^{-4}} \right]
\left. \right\}
\label{xzero}
\eea 
The third line of Eq. (\ref{xzero}) describes the exchange  of
 the lightest (massless when $a_i^{I_1I_3}=0$) 33 states  as well as
 their winding  excitations with respect to the four  ND
 directions. The numerator is a form factor multiplying the coupling
 of a 33 state to two 37 states which originates from the interaction
 of two twisted fields to an untwisted winding mode \cite{twist,AB,narain}.
 Similarly, the second line of Eq. (\ref{xzero}) describes the
 exchange  of the massless 77 states together with their KK
 excitations. Notice the presence of the $1/R_i$ prefactor which
 reflects the  suppression of the coupling of the D7 brane  states
 compared to those of the D3 branes  by the volume of the
 extra dimensions (bulk).

The exponential form factor of the coupling of KK excitations can be
understood from the behavior of D3 branes  as solitonic objects with
finite thickness inside D7 branes. In fact the interaction of the KK
excitations of the D7 gauge fields $A^\mu (x,\vec y)=\sum_{{\vec n}}
\A^{\mu }_{{\vec n}} \exp{i\frac {n_i y_i}{R_i}}$ with the charge
density $j_\mu (x)$ associated to massless  37 fermions is described
by the effective Lagrangian: 
\bea 
\int d^4x \, \, \,  \, \sum_{{\vec
n}} e^{-\ln {\delta} \sum_{i=1}^{4}\frac{n_i^2l_s^2}{2 R_i^2}} \, \,
\, \, \, j_\mu (x) \, \A^{\mu }_{\vec n}(x)\, , \eea 
 which can be written after Fourier transform as 
\be \int d^{4}y \,\int d^4x  \, \,
\, \,  (\frac{1}{l_s^2 2 \pi \ln {\delta}})^{2} e^{- \frac {{\vec
y}^2}{2 l_s^2 \ln {\delta}}}  \, j_\mu (x) \, A^\mu (x,\vec y)\, .
\label{brwidth}
\ee
It follows that the D3 brane appears from the viewpoint of the D7 brane as a Gaussian
distribution of charge $e^{-\frac {{\vec y}^2}{2 \sigma^2}} j_\mu (x)$  with a width
$\sigma=\sqrt{\ln {\delta}}\, l_s \sim 1.66 \, l_s$.
Note the similarity of this result with the numerical estimate of the
D-brane width arising from the analysis of tachyon condensation, leading to
$\sigma\sim 1.55\,  l_s$ \cite{szwie}.

The limit $x \rightarrow 1$ of eq. (\ref{mastereq}) can  be obtained from 
(\ref{xzero}) by exchanging D7 brane indices with the D3 brane ones and 
$s$ with $t$: 
\bea
{\lim _{x\rightarrow 1}}\A (1_{j_1 I_1},2_{j_2 I_2},3_{j_3 I_3},4_{j_4 I_4})
&= &  
 {g_s} \, \,   \left[ {\bar u}^{(1)} \gamma_{\mu} u^{(4)} 
{\bar u}^{(2)} \gamma^{\mu} u^{(3)}   \right ] \, \,\nonumber \\ 
 \times \{  \delta_{I_1,{\bar I_2}} \delta_{I_3,{\bar I_4}}
\delta_{{\bar j_1},j_4} \delta_{j_2,{\bar j_3}}   
&& \left[ \sum_{n_i \in {\bf Z}} \frac 
{\delta^{-   \sum_{i=1}^{4} (n_i +a_i^{I_1I_3})^2\,  R_i^2 \, l_s^{-2} }}
{t  -   \sum_{i=1}^{4} (n_i +a_i^{I_1I_3})^2\,  R_i^2 \, l_s^{-4}} \right]
\nonumber \\ +
\frac { \delta_{j_1,{\bar j_2}} 
\delta_{j_3,{\bar j_4}}
\delta_{{\bar I_1}, I_4} \delta_{I_2,{\bar I_3}}  }
{ l_s^{-4}\, \prod_{i=1}^{4} R_i}  
&& \left[ \frac {1}{t} +\sum_{m_i \in {\bf Z}-\{ 0\}} \frac {e^{i  2 \pi \sum_{i=1}^{4}
 m_i a_i^{I_1I_3}} \, \, \delta^{  - \sum_{i=1}^{4} \frac{m_i^2 l_s^2}{ R_i^2}}}
{t - \sum_{i=1}^{4} 
\frac{m_i^2}{ R_i^2}} \right] \}  \nonumber \\
\label{tired}
\eea

In the following, we will restrict to the case of maximal gauge
symmetry  $a_i^{I}=0$ and  we  choose all radii to be bigger than the
string scale, $R_i> l_s$. The contribution of massless 33 and 77
states as well as the contribution of KK excitations of the latter
are computable in the effective field theory. The  
amplitude  receives two types of stringy
contributions originating from the exchange of  either string winding
modes or string oscillators. At low energies, below the string scale,
both winding and oscillator states are heavy and can be integrated out
to generate effective contact interactions. Their general form is
\bea 
\A^{cont} = \A^{cont}_{w} + \A^{cont}_{osc}\, ,
\label{defcon}
\eea
where from Eqs. (\ref{xzero}) and (\ref{tired}) 
\bea
\A^{cont}_{w} &=& g_s l_s^2  \left\{  \right.
\delta_{I_1,{\bar I_2}} \delta_{I_3,{\bar I_4}}
\delta_{{\bar j_1},j_4} \delta_{j_2,{\bar j_3}}
\left[{\bar u}^{(1)} \gamma_{\mu} u^{(4)} 
{\bar u}^{(2)} \gamma^{\mu} u^{(3)}\right]   
\nonumber \\ &+&\!  
\delta_{j_1,{\bar j_2}} 
\delta_{j_3,{\bar j_4}}
\delta_{{\bar I_1}, I_4} \delta_{I_2,{\bar I_3}}
\left[  {\bar u}^{(1)} \gamma_{\mu} u^{(2)} 
{\bar u}^{(4)} \gamma^{\mu} u^{(3)}\right] \left. \! \right\} 
\sum_{n_i \in {\bf Z}-\{ 0\}} \frac 
{\delta^{-   \sum_{i=1}^{4} n_i^2\,  R_i^2 \, l_s^{-2} }}
{  \sum_{i=1}^{4} n_i^2\,  R_i^2 \, l_s^{-2}}
\label{aw}
\eea
On the other hand,  
\bea
\A^{cont}_{osc} = {\lim _{sl_s^2,tl_s^2\rightarrow 0}} 
\{\A - \A^{QFT}\} \quad {\rm with} \quad
\A^{QFT} =({\lim _{x\rightarrow 0}} +{\lim _{x\rightarrow 1}})\A\, ,
\label{defQFT}
\eea
where $\A^{QFT}$ contains also the contribution of 33 string winding modes, in
addition to the field theory part from exchanges of massless modes and 77 KK
excitations.

\subsection {Four ND fermions from  three sets of branes}

We discuss now  the case where the fermions
arise from open strings lying on different sets of D7 branes.
 Without loss of generality, the latter can be chosen to be the 
second and  third sets of table 1, transverse to the 6,7 and
8,9 directions, respectively.

The computation goes as before with the difference that 
it involves correlations of four twist fields only in the 4th and 5th 
coordinates while in the remaining four internal directions it involves 
correlations of two twist fields. The result for the ordered amplitude is 

\bea
 \A (1_{j_1 I_1},2_{j_2 I_2},3_{j_3 I_3},4_{j_4 I_4})&=&   
-  g_s \, l_s^2 \int_0^1 dx \, \, x^{-1 -s\, l_s^2}\, \, \,  
(1-x)^{-1 -t\, l_s^2} \, \, \,  \frac {1}{ F (x) }\, \, \,  \nonumber \\  
\{ \delta_{I_1,{\bar I_2}} \delta_{I_3,{\bar I_4}}
 \delta_{{\bar j_1}, j_4} \delta_{j_2,{\bar j_3}}  
&&  \left[ \right.
 \frac {  {\bar u}^{(1)} u^{(2)}  {\bar u}^{(4)}  u^{(3)} \, (1-x)} 
{ l_s^{-2} \prod_{i=1}^{2} R_i} 
\, \sum_{m_i\in {\bf Z}} e^{2 i \pi \sum_{i=1}^{2} m_i a_i^{I_1I_3}} \, \, 
e^{ - {\pi} {\tau}\, \sum_i \frac {  m_i ^2 \, l_s^2 }{ R_i^2} } 
\nonumber \\ 
&& + {\bar u}^{(1)} \gamma^{\mu} u^{(4)}  {\bar u}^{(2)}\gamma_{\mu} u^{(3)}\, x \,
\sum_{n_i\in {\bf Z}}  e^{-  {\pi \tau}  \sum_{i=1}^{2} (n_i +a_i^{I_1I_3})^2\,  R_i^2 \, l_s^{-2} }\left. \right]
\nonumber \\ 
+ \delta_{j_1,{\bar j_2}} 
\delta_{j_3,{\bar j_4}}
\delta_{{\bar I_1}, I_4} \delta_{I_2,{\bar I_3}}   
&&  \left[ \right.
 \frac {  {\bar u}^{(1)} u^{(4)}  {\bar u}^{(2)}  u^{(3)} \, x} 
{ l_s^{-2} \prod_{i=1}^{2} R_i} 
\, \sum_{m_i\in {\bf Z}} e^{2 i \pi \sum_{i=1}^{2} m_i a_i^{I_1I_3}} \, \, 
e^{ - {\pi} {\tau}\, \sum_i \frac {  m_i ^2 \, l_s^2 }{ R_i^2} } 
\nonumber \\ 
&& + {\bar u}^{(1)} \gamma^{\mu} u^{(2)}  {\bar u}^{(4)}\gamma_{\mu} u^{(3)}\, (1-x) \,
\sum_{n_i\in {\bf Z}}  e^{-  {\pi \tau}  \sum_{i=1}^{2} (n_i +a_i^{I_1I_3})^2\,  R_i^2 \, l_s^{-2} }\left. \right] \,  \}
\nonumber \\ 
\label{twot}
\eea
Taking the limit of coincident
vertices  $x \rightarrow 0$ or $x \rightarrow 1$,  one obtains the 
field theory result which for  $a_i^{I}=0$ reads:
\bea 
 \A^{QFT} =  g_s \, 
 \delta_{I_1,{\bar I_2}} \delta_{I_3,{\bar I_4}}
\delta_{{\bar j_1}, j_4} \delta_{j_2,{\bar j_3}}  
\,   &\{&  \frac {  \left[ {\bar
u}^{(1)} u^{(2)}  {\bar u}^{(4)}  u^{(3)}
\right ]} { l_s^{-2} \prod_{i=1}^{2} R_i} 
\,
\left[ \frac {1}{s} +\sum_{m_i \in {\bf Z}-\{ 0\}} 
\frac { \delta^{  - \sum_{i=1}^{2} \frac{m_i^2 l_s^2}{ R_i^2}}}
{s - \sum_{i=1}^{2} 
\frac{m_i^2}{ R_i^2}} \right] \nonumber \\ 
+ 
\left[   {\bar u}^{(1)} \gamma^{\mu} u^{(4)}  {\bar u}^{(2)}\gamma_{\mu} u^{(3)} \right ] &&
\left[\frac {1}{t} +\sum_{n_i\in {\bf Z}-\{ 0\}} \frac 
{\delta^{-   \sum_{i=1}^{2} n_i^2\,  R_i^2 \, l_s^{-2} }}
{t  -   \sum_{i=1}^{2} n_i^2\,  R_i^2 \, l_s^{-4}} \right]
 \, \,  \, \, 
\} \nonumber \\
+ g_s \, 
\delta_{j_1,{\bar j_2}} 
\delta_{j_3,{\bar j_4}}
\delta_{{\bar I_1}, I_4} \delta_{I_2,{\bar I_3}}  \,    &\{&   \frac{\left[   {\bar u}^{(1)}  u^{(4)}  {\bar u}^{(2)} u^{(3)} \right ]} 
{ l_s^{-2}\,
\prod_{i=1}^{2} R_i}  
\left[ \frac {1}{t} +\sum_{m_i \in {\bf Z}-\{ 0\}} 
\frac { 
\delta^{  - \sum_{i=1}^{2} \frac{m_i^2 l_s^2}{ R_i^2}}}
{t - \sum_{i=1}^{2} \frac{m_i^2}{ R_i^2}} \right] 
\nonumber \\+ 
\left[ {\bar u}^{(1)} \gamma_{\mu} u^{(2)}  {\bar u}^{(4)} 
\gamma^{\mu} u^{(3)}
\right ] &&
\left[ \frac {1}{s} +\sum_{n_i \in {\bf Z}-\{ 0\}} \frac 
{\delta^{-   \sum_{i=1}^{2} n_i^2\,  R_i^2 \, l_s^{-2} }}
{s  -   \sum_{i=1}^{2} n_i^2\,  R_i^2 \, l_s^{-4}} \right]
\, \,  \, \, \} .
\eea
Note that only the KK excitations and the winding modes associated with the two
internal dimensions longitudinal to both D7 branes are exchanged among the 
four fermions. Moreover, while the exchanged 33 states are vector bosons, 
the exchanged 77' states are spin 0 particles, thus the absence of 
$\gamma_{\mu}$-matrices.

\subsection {Two ND and two DD fermions }

For completeness we also present the last possibility where 
only two of the fermions arise from ND open strings stretched between 
the same set of branes. The computation is straightforward \cite{hash} 
and gives a result similar to the case of four DD fermions:
\bea
\A(1_{j_1 I_1},2_{j_2 I_2},3,4)& =& -2 g_s \, l_s^2  \, 
\{ \lambda^3, \lambda^4 \}_{I_1{\bar I_2}}  \delta_{j_1,{\bar j_2}} 
 \, \,
\int_0^1 dx \, \,  x^{-1 -s\, l_s^2} \,  \, (1-x)^{-1 -t\, l_s^2} \, \,  
\nonumber \\
&\times& [   {\bar u}^{(1)} \gamma_{\mu} u^{(2)} 
{\bar u}^{(4)} \gamma^{\mu} u^{(3)}   (1-x)  +
 {\bar u}^{(1)}  u^{(4)}  
{\bar u}^{(2)} u^{(3)} x  ]\, , \nonumber
\label{noknown}
\eea
where the absence of $\gamma$-matrices in the $t$ channel is due to the fact 
that it corresponds to the exchange of 37 open strings which do not contain
spin-one modes. 
The case of two ND and two DD fermions is thus very similar to the one of 
four DD strings given in eq. (\ref{known}); the effective action does not 
contain dimension-six contact interactions but the leading correction comes
from dimension-eight operators.

\section {Effective four-fermion contact interactions}

In this section we would like to compute the contact interaction
$\A^{contact}$ between four fermions induced by massive string states:
oscillators and  winding modes.  We will first study
the case of four ND fermions from  two sets of branes where the relevant 
amplitudes have been defined in Eqs.  (\ref{defcon}) to  (\ref{defQFT}).
The other cases will be discussed in the subsection 4.3.

\subsection {Two sets of branes: the infinite transverse volume case}

We first consider the simplest case  obtained by taking the size of the
directions appearing in the result of the correlation functions 
(\ref{fourtw}) of twist fields  to be very large: $R_i/ l_s \to \infty$. 
In this limit the contribution of winding modes $\A^{cont}_{w}$
given by (\ref{aw}) 
vanishes and the contact interaction is due only to massive string 
oscillation modes 
$\A^{cont} \simeq \A^{cont}_{osc}$ given in Eq. (\ref{defQFT}).
The field theory part $\A^{QFT}$ is due to 
the exchange of massless 33 modes:
\bea
 \A^{33} = 
\delta_{j_1,{\bar j_2}} \delta_{j_3,{\bar j_4}}
\delta_{{\bar I_1}, J_4} \delta_{I_2,{\bar I_3}} 
  &&
 \left[   {\bar u}^{(1)} \gamma_{\mu} u^{(2)}  {\bar u}^{(4)}
\gamma^{\mu} u^{(3)} \right ] \, \, \frac {g_s} { s }    
 \nonumber \\
+ \delta_{I_1,{\bar I_2}} \delta_{I_3,{\bar I_4}}
\delta_{{\bar j_1}, j_4} \delta_{j_2,{\bar j_3}}    &&
 \left[ {\bar u}^{(1)} \gamma_{\mu} u^{(4)} 
{\bar u}^{(2)} \gamma^{\mu} u^{(3)}   \right ]  \, \, 
\frac {g_s} { t }\, ,
\label{spol}
\eea
as well as to the exchange of massless 77 states and their KK excitations:
\bea 
 \A^{77} =  \delta_{j_1,{\bar j_2}} 
\delta_{j_3,{\bar j_4}}
\delta_{{\bar I_1}, I_4} \delta_{I_2,{\bar I_3}}
 &&  \frac {  \left[ {\bar
u}^{(1)} \gamma_{\mu} u^{(4)}  {\bar u}^{(2)} \gamma^{\mu} u^{(3)}
\right ]} {  \prod_{i} (R_i/l_s)} 
\,  \sum_{m_i } \frac
{ g_s  \, \, \delta^{  - \sum_i \frac{m_i^2
l_s^2}{ R_i^2}}}{t - \sum_i \frac{m_i^2}{ R_i^2}} 
 \nonumber \\ 
+ \delta_{I_1,{\bar I_2}} \delta_{I_3,{\bar I_4}}
\delta_{{\bar j_1}, j_4} \delta_{j_2,{\bar j_3}}  
   &&   \frac{\left[   
{\bar u}^{(1)} \gamma_{\mu} u^{(2)}  {\bar u}^{(4)}
\gamma^{\mu} u^{(3)} \right ]} { 
\prod_{i} (R_i/l_s)}  \sum_{m_i} \frac
{ g_s \, \, \delta^{  - \sum_i \frac{m_i^2 l_s^2}{ R_i^2}}}{s  - 
\sum_i \frac{m_i^2}{ R_i^2}} 
\label{tpol}
\eea
where $i= 1,\cdots,4$.

The contact interaction is then given by:
\bea
 \A^{contact}|_{R_i/ l_s \to \infty} =\A|_{R_i/ l_s \to \infty} - 
 \A^{33}- \A^{77}
\label{crinf} 
\eea
Notice that each term in the r.h.s. of Eq. (\ref{crinf}) has two 
contributions proportional to $\delta_{j_1,{\bar j_2}}
\delta_{j_3,{\bar j_4}} \delta_{{\bar I_1}, I_4} \delta_{I_2,{\bar
I_3}}$ and to
$\delta_{I_1,{\bar I_2}} \delta_{I_3,{\bar I_4}} \delta_{{\bar j_1},
j_4} \delta_{j_2,{\bar j_3}}$. These contributions are 
related by the exchange of the states 2 and 4, which takes $s$ to $t$.
To verify this property in the integral representation of the 
amplitude (\ref{mastereq})  one should  change $x$ 
to $1-x$ and perform a Poisson ressumation. We can thus restrict our analysis
to the terms proportional to $\delta_{j_1,{\bar j_2}}
\delta_{j_3,{\bar j_4}} \delta_{{\bar I_1}, I_4} \delta_{I_2,{\bar
I_3}}$.

The ordered amplitude $\A$ in (\ref{mastereq})  
contains  the factor $\sum_{n_i}  e^{-  {\pi \tau}  \sum_i n_i^2\,  R_i^2
\, l_s^{-2} }$ which vanishes exponentially in 
the limit ${R_i/ l_s \to \infty}$ unless $n_i=0$ or $\tau \to 0$.
The limit $\tau \to 0$ corresponds  to $x \to 1$, 
as it can be seen from 
(\ref{limtau}), and produces the field theory result 
$\A^{77}$ of Eq. (\ref{tpol}) which we must  subtract according to Eq. (\ref{crinf}).
We are left over with the 
contribution of the zero mode $n_i=0$, given by the integral 
(\ref{mastereq}) after substracting the integration region $x \to 1$ :
\bea 
- { g_s} \, l_s^2\, \int_0^1\! &dx& \left\{ \right.
x^{-1 -s\, l_s^2}   (1-x)^{-1 -t\, l_s^2}  \! \frac {
\left[   {\bar u}^{(1)}\! \gamma_{\mu} u^{(2)}\!  {\bar u}^{(4)}\!
\gamma^{\mu} u^{(3)}\! (1-x) + {\bar u}^{(1)}\! \gamma_{\mu}\!  u^{(4)}
{\bar u}^{(2)}\! \gamma^{\mu}\! u^{(3)}  x \right ]}{ [F (x)]^{2} }\nonumber \\
&-&  \, \, (1-x)^{-1 -t\, l_s^2} \, \, \, 
\left[{\bar u}^{(1)} \gamma_{\mu} 
u^{(4)} {\bar u}^{(2)} \gamma^{\mu} u^{(3)} \right ]
 [\frac {\pi} {\ln {\frac {\delta}{1- x}}}]^2 \left. \right\}
\label{part1d3}
\eea
This contains a simple pole in $s$ due to the exchange of massless
33 strings given in Eq. (\ref{spol}). Following Eq. (\ref{crinf}) this
pole should also  be substracted by adding to (\ref{part1d3}) the term
\bea
{ g_s} \, l_s^2 \, \int_0^1 dx \, \, x^{-1 -s\, l_s^2}\, \, \,   \left[   
{\bar u}^{(1)} \gamma_{\mu} u^{(2)} 
{\bar u}^{(4)} \gamma^{\mu} u^{(3)}\right ]\, .
\label{part2d3}
\eea

Adding (\ref{part1d3}) and (\ref{part2d3}) and taking the low energy limit 
$|s \, l_s^2| \to 0$ and  $|t\,  l_s^2| \to 0$ we obtain the 
dimension-six effective operator:
\bea
- g_s l_s^2 \left[   {\bar u}^{(1)} \gamma_{\mu} u^{(2)} 
{\bar u}^{(4)} \gamma^{\mu} u^{(3)}\right ] && \int_0^1 \frac {dx}{x} ( 
\frac {1}{ [F (x)]^2 } -1 )\nonumber \\
- g_s l_s^2 \left[{\bar u}^{(1)} \gamma_{\mu} 
u^{(4)} {\bar u}^{(2)} \gamma^{\mu} u^{(3)} \right ] && \int_0^1 \frac {dx}
{1-x} (\frac {1}{ [F (x)]^2 } - [\frac {\pi} {\ln {\frac {\delta}{1- x}}}]^2)
\label{xrinf}
\eea 

To give a numerical evaluation of the above four-fermion interaction,
it is convenient to make a change of variables from $x$ to $\tau$
using Eq. (\ref{xtotau}). The integration domain  is then divided into
two regions: $\tau \in [0,1]$ and  $\tau \in [1,\infty]$.  To make the
convergence of the integrals manifest we perform a transformation
$\tau \to 1/\tau$ in the region $\tau \in [0,1]$ so that the
integration is now only from 1 to $\infty$. 
Adding the  two 
contributions proportional to $\delta_{j_1,{\bar j_2}}
\delta_{j_3,{\bar j_4}} \delta_{{\bar I_1}, I_4} \delta_{I_2,{\bar
I_3}}$ and 
$\delta_{I_1,{\bar I_2}} \delta_{I_3,{\bar I_4}} \delta_{{\bar j_1},
j_4} \delta_{j_2,{\bar j_3}}$, we obtain for the  four-fermion
contact term the following result:
\bea
 \A^{contact} &=& - \delta_{j_1,{\bar j_2}} 
\delta_{j_3,{\bar j_4}}
\delta_{{\bar I_1}, I_4} \delta_{I_2,{\bar I_3}}  
\, \,\, \,  \, \,   {g_s}  \, l_s^2  \, \{\,  \, \,   \, \,  
\left[ {\bar u}^{(1)} \gamma_{\mu} u^{(4)}  {\bar u}^{(2)} 
\gamma^{\mu} u^{(3)} \right ] \nonumber \\ &\times & 
\pi \int_1^\infty d\tau \, (\frac {\theta_2^4(i \tau) }{ \theta_3^4(i \tau)}
+ \frac {1}{\tau^2} \frac {\theta_4^4(i \tau) }{\theta_3^4(i \tau)}
- {\theta_2^4(i \tau) }  [\frac {\pi}{4 \ln{(2 \frac {\theta_3}{\theta_4})}}]^2
- {\theta_4^4(i \tau) }  [\frac {\pi}{4 \ln{(2 \frac {\theta_3}{\theta_2})}}]^2 )\nonumber \\ 
&+& 
 \left[   {\bar u}^{(1)} \gamma_{\mu} u^{(2)}  {\bar u}^{(4)}
\gamma^{\mu} u^{(3)} \right ]  \pi \int_1^\infty d\tau \, ( 
\frac {\theta_4^4(i \tau) }{\theta_3^4(i \tau)} - {\theta_3^4(i \tau)} +
\frac {1}{\tau^2}\frac {\theta_2^4(i \tau) }{ \theta_3^4(i \tau)} )
 \, \,  \, \, 
\} \nonumber \\
&-& 
 \delta_{I_1,{\bar I_2}} \delta_{I_3,{\bar I_4}}
\delta_{{\bar j_1}, j_4} \delta_{j_2,{\bar j_3}}  
\, \,\, \,  \, \,  {g_s}  \, l_s^2  \, \{ \,  \, \,   \, \
\left[   {\bar u}^{(1)} \gamma_{\mu} u^{(2)}  {\bar u}^{(4)}
\gamma^{\mu} u^{(3)} \right ] \nonumber \\ &\times & 
\pi \int_1^\infty d\tau \, (\frac {\theta_2^4(i \tau) }{ \theta_3^4(i \tau)}
+ \frac {1}{\tau^2} \frac {\theta_4^4(i \tau) }{\theta_3^4(i \tau)}
- {\theta_2^4(i \tau) }  [\frac {\pi}{4 \ln{(2 \frac {\theta_3}{\theta_4})}}]^2
- {\theta_4^4(i \tau) }  [\frac {\pi}{4 \ln{(2 \frac {\theta_3}{\theta_2})}}]^2 )
 \nonumber \\ 
&+& 
 \left[ {\bar u}^{(1)} \gamma_{\mu} u^{(4)} 
{\bar u}^{(2)} \gamma^{\mu} u^{(3)}   \right ]
  \pi \int_1^\infty d\tau \, ( 
\frac {\theta_4^4(i \tau) }{\theta_3^4(i \tau)} - {\theta_3^4(i \tau)} +
\frac {1}{\tau^2}\frac {\theta_2^4(i \tau) }{ \theta_3^4(i \tau)} )
 \, \,  \, \, 
\}
\eea
which can be approximated by numerical integration as:
 \bea 
 \A^{contact}\! \simeq\! &&\! \! {g_s} l_s^2
\delta_{j_1,{\bar j_2}} 
\delta_{j_3,{\bar j_4}}
\delta_{{\bar I_1}, I_4} \delta_{I_2,{\bar I_3}}  
 \{ 0.20
 {\bar u}^{(1)} \gamma_{\mu} u^{(4)}  {\bar u}^{(2)} 
\gamma^{\mu} u^{(3)} \! + 0.59
  {\bar u}^{(1)} \gamma_{\mu} u^{(2)}  {\bar u}^{(4)}
\gamma^{\mu} u^{(3)}    
\} \nonumber \\
&+& \! \!   g_s l_s^2
 \delta_{I_1,{\bar I_2}} \delta_{I_3,{\bar I_4}}
\delta_{{\bar j_1}, j_4} \delta_{j_2,{\bar j_3}}  
 \{  
0.59 {\bar u}^{(1)} \gamma_{\mu} u^{(4)}  {\bar u}^{(2)} 
\gamma^{\mu} u^{(3)} \! + 0.20
  {\bar u}^{(1)} \gamma_{\mu} u^{(2)}  {\bar u}^{(4)}
\gamma^{\mu} u^{(3)}   
\}\nonumber \\
\label{approx}
\eea
In eq. (\ref{approx}), the leading contribution (with coefficient $0.59$)
arises from exchange of massive oscillators of 33 states and will be used to extract experimental bounds on the string scale in section 6.

Note that the coefficients of the above contact operators are positive i.e. 
opposite to 
those due to the exchange of massive KK excitations of gauge bosons.
The non-vanishing effective operators  for fermions of the same type (all 
the D3 and D7 brane indices equal $\delta_{I_a,{\bar I_b}}=\delta_{{\bar j_a}, j_b} =1$)
with fixed helicities are:
\bea 
\A(1_L^-, 2_R^+ , 3_L^-, 4_R^+)&\simeq&  3.16  \, g_s \, {u}{l_s^2}\, , \\ 
\A(1_L^-, 2_R^+ , 3_R^-, 4_L^+)&\simeq& - 1.58  \, g_s \, {t}{l_s^2}\, , \\
\A(1_L^-, 2_L^+ , 3_L^- ,4_L^+)&\simeq& - 1.58  \, g_s \, {s}{l_s^2}\, ,  
\label{ac}
\eea
while the other non-vanishing operators can be obtained by parity reflection.

\subsection {Two sets of branes: generalization to arbitrary internal radii}

We consider now the generic case where the sizes of
 $d$ directions associated with the KK and winding modes exchanged 
between the four fermions are kept finite while the remaining are taken 
infinitely large. Following the definition of $\A^{contact}$ in
(\ref{defcon}), we substract the ${\cal A}^{33}$ and ${\cal A}^{77}$
contributions from the total ordered amplitude; this gives 
  for the term proportional to $\delta_{I_1,{\bar I_2}} \delta_{I_3,{\bar I_4}}
\delta_{{\bar j_1}, j_4} \delta_{j_2,{\bar j_3}}$:
\bea
\A^{contact}&=& - g_s\, l_s^2\, [{\bar u}^{(1)} \gamma_{\mu} u^{(2)} 
{\bar u}^{(4)} \gamma^{\mu} u^{(3)}]\int_0^1 \frac {dx}{x} ( 
\frac {1}{[F (x)]^2}\sum_{n_i}
e^{ - {\pi} {\tau}\sum_{i=1}^{d}\frac{n_i^2R^2}{l_s^2}}-1)\nonumber \\
&-& g_s\, l_s^2\, [{\bar u}^{(1)} \gamma_{\mu} 
u^{(4)} {\bar u}^{(2)} \gamma^{\mu} u^{(3)}]
\int_0^1 \frac {dx}
{1-x} \nonumber \\
&&(\frac {1}{ [F (x)]^2 }\sum_{n_i}
e^{ - {\pi} {\tau}\sum_{i=1}^{d}\frac{n_i^2R_i^2}{l_s^2}} -
\frac{l_s^{d}}{\prod_{i=1}^{d}R_i}
[\frac{\pi}{\ln{(\frac{\delta}{1-x})}}]^{\frac{4-d}{2}}\sum_{n_i}
[\frac{(1-x)}{\delta}]^{\sum_{i=1}^{d}\frac{n_i^2l_s^2}{R_i^2}})\nonumber \\
\label{cond}
\eea
where in the last term inside the brackets, corresponding to the  
$x \to 1$ asymptotic region, we performed a Poisson resummation.
The above contact term contains contributions of both string oscillators and 
winding states along the finite size directions, in contrast to the 
large radius limit for all directions that we considered in subsection 4.1. 
The contribution of the winding modes reads:
\bea
\A^{cont}_{w} &=& - g_s \, l_s^2\, 
\left[  {\bar u}^{(1)} \gamma_{\mu} u^{(2)} 
{\bar u}^{(4)} \gamma^{\mu} u^{(3)}\right] 
\sum_{n_i \in {\bf Z}-\{ 0\}} \frac 
{\delta^{-   \sum_{i=1}^{d} n_i^2\,  R_i^2 \, l_s^{-2} }}
{  \sum_{i=1}^{d} n_i^2\,  R_i^2 \, l_s^{-2}}
\label{wcon}
\eea 

To study the dependance of the  contact term (\ref{cond}) 
on the size of the compactification space
we consider the simplest case of one finite size dimension and three infinite ones ($d=1$).
As in the case of infinite radius, for the numerical evaluations 
 it is useful to express  the corresponding  integral 
as a function of $\tau$. The resulting expression is:
\bea
\A^{contact}(R) &=&- {g_s} \, l_s^2 \qquad
\delta_{I_1,{\bar I_2}} \delta_{J_3,{\bar J_4}}
\delta_{{\bar j_1}, j_4} \delta_{j_2,{\bar j_3}}  \nonumber\\
&& [ [{\bar u}^{(1)} \gamma_{\mu} u^{(4)}  {\bar u}^{(2)} 
\gamma^{\mu} u^{(3)}]  \pi 
\int_1^\infty d\tau 
(\frac {\theta_2^4 }{
 \theta_3^4}(i\tau)\theta_3(i\frac{R^2\tau}{l_s^2})+\frac {l_s}{R\tau^\frac{3}{2}}
 \frac {\theta_4^4
 }{\theta_3^4}(i\tau)\theta_3(i\frac{l_s^2\tau}{R^2}) \nonumber \\
&-& {\theta_2^4(i \tau)
 }\frac{l_s}{R}  [\frac {\pi}{4 \ln{(2 \frac
 {\theta_3}{\theta_4}(i\tau))}}]^\frac{3}{2}\sum_{n \in Z}[\frac{\theta_4}{2\theta_3}(i\tau)]^\frac{4n^2l_s^2}{R^2}\nonumber \\
&-& {\theta_4^4(i \tau) }\frac{l_s}{R}  [\frac {\pi}{4 \ln{(2 \frac
    {\theta_3}{\theta_2}(i\tau))}}]^\frac{3}{2}\sum_{n \in Z}[\frac{\theta_2}{2\theta_3}(i\tau)]^\frac{4n^2l_s^2}{R^2} )\nonumber  \\
&+&\left[   {\bar u}^{(1)} \gamma_{\mu} u^{(2)}  {\bar u}^{(4)}
\gamma^{\mu} u^{(3)} \right ] \pi \int_1^\infty d\tau \, ( 
\frac {\theta_4^4}{\theta_3^4}(i \tau)\theta_3(i\frac{R^2\tau}{l_s^2}) - {\theta_3^4(i \tau)} 
\nonumber \\ &+&
\frac {l_s}{R\tau^\frac{3}{2}}\frac {\theta_2^4}{ \theta_3^4}(i \tau)\theta_3(i\frac{l_s^2\tau}{R^2}) )]
\nonumber  \\ &-&{g_s}\, l_s^2 \delta_{j_1,{\bar j_2}} 
\delta_{j_3,{\bar j_4}}
\delta_{{\bar I_1}, J_4} \delta_{I_2,{\bar J_3}} \left\{ \right.
[{\bar u}^{(1)} 
\gamma_{\mu} u^{(2)}  {\bar u}^{(4)}
\gamma^{\mu} u^{(3)} ]  \pi\int_1^\infty d\tau \nonumber \\
&&(\ \ \frac {\theta_2^4 }{
 \theta_3^4}(i\tau)\theta_3(i\frac{R^2\tau}{l_s^2})
+ \frac {l_s}{R\tau^\frac{3}{2}}
 \frac {\theta_4^4
 }{\theta_3^4}(i\tau)\theta_3(i\frac{l_s^2\tau}{R^2})
-{\theta_2^4(i \tau)
 }\frac{l_s}{R}  [\frac {\pi}{4 \ln{(2 \frac
 {\theta_3}{\theta_4}(i\tau))}}]^\frac{3}{2}\nonumber \\
&& \sum_{n \in Z}
[\frac{\theta_4}{2\theta_3}(i\tau)]^\frac{4n^2l_s^2}{R^2}
-{\theta_4^4(i \tau) }\frac{l_s}{R}  
[\frac {\pi}{4 \ln{(2 \frac
   {\theta_3}{\theta_2}(i\tau))}}]^\frac{3}{2}\sum_{n \in
Z}[\frac{\theta_2}{2\theta_3}(i\tau)]^\frac{4n^2l_s^2}{R^2} \ \ ) 
 \nonumber \\&+&[{\bar u}^{(1)} \gamma_{\mu} u^{(4)} 
{\bar u}^{(2)} \gamma^{\mu} u^{(3)}]
  \pi\int_1^\infty d\tau  (\ \ 
\frac {\theta_4^4}{\theta_3^4}(i\tau)\theta_3(i\frac{R^2\tau}{l_s^2}) - {\theta_3^4(i
\tau)}\nonumber \\ && \qquad\qquad\qquad\qquad\qquad\qquad  +
\frac {l_s}{R\tau^\frac{3}{2}}\frac {\theta_2^4}{
\theta_3^4}(i\tau)\theta_3(i\frac{l_s^2\tau}{R^2})\ \ )\left. \right\} 
\label{unR}
\eea 
The behavior as a function of $R/ l_s$ of the contact term due to massive 33 string
oscillator states (given by the first line of eq. (\ref{cond}) or equivalently
by the integral in the 5th and 6th lines of eq. (\ref{unR}) ) is shown in
figure 1 (solid line).
\begin{figure}[h]
\centering
\epsfysize=4.5in
\includegraphics[width=9cm,height=6cm]{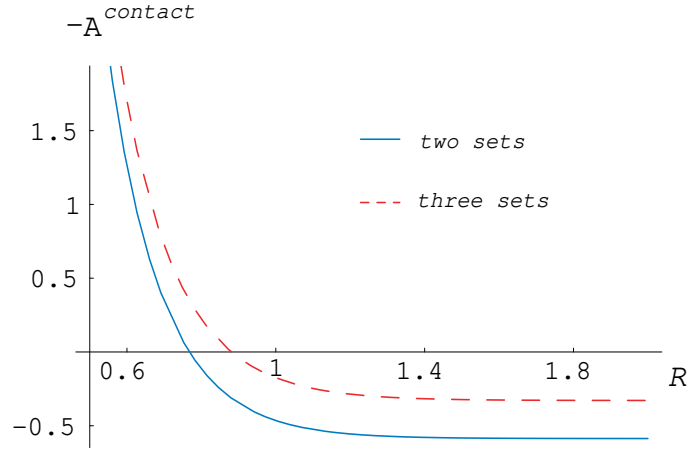}
\caption{Radius dependence of the four-fermion contact terms for two (solid line) and
three (dashed line) sets of branes.}
\label{effpot}
\end{figure}
In the limit $R/ l_s\to\infty$, it reaches rapidly its asymptotic value 0.59 of eq.
(\ref{approx}), while when $R/ l_s\to 0$, it diverges as $-\pi^2l_s^2/3R^2$,
obtained from the sum (\ref{wcon}) with $d=1$. This limit corresponds, upon
T-duality $R\to l_s^2/R$, to the decompactification of a longitudinal dimension
along the world-volume of a D4 brane.

\subsection {Three sets of branes} 

In this case, there are two different sets of 7-branes and the corresponding amplitude is
given in eq. (\ref{twot}). The difference with the previous case is that there are two
(instead of four) internal directions that appear in the amplitude and that the 77' channel
leads only to scalar exchanges. For $d$ arbitrary internal radii ($d\le 2$), the
contribution  proportional to $\delta_{j_1,{\bar j_2}}
\delta_{j_3,{\bar j_4}} \delta_{{\bar I_1}, I_4} \delta_{I_2,{\bar
I_3}}$, previously given in eq. (\ref{cond}),
becomes:
\bea
\A^{contact}&=& - g_s\, l_s^2\, [{\bar u}^{(1)}\gamma_{\mu} u^{(2)} {\bar u}^{(4)}\gamma^{\mu} u^{(3)}] 
\int_0^1 \frac {dx}{x} ( 
\frac {1}{F (x)}\sum_{n_i}
e^{ - {\pi} {\tau}\sum_{i=1}^{d}\frac{n_i^2R^2}{l_s^2}}-1)\nonumber \\
&-& g_s\, l_s^2\, [{\bar u}^{(1)} 
u^{(4)} {\bar u}^{(2)}  u^{(3)}]
\int_0^1 \frac {dx}
{1-x} \nonumber \\
&&(\frac {1}{ F (x)}\sum_{n_i}
e^{ - {\pi} {\tau}\sum_{i=1}^{d}\frac{n_i^2R_i^2}{l_s^2}} -
\frac{l_s^{d}}{\prod_{i=1}^{d}R_i}
[\frac{\pi}{\ln{(\frac{\delta}{1-x})}}]^{\frac{2-d}{2}}\sum_{n_i}
[\frac{(1-x)}{\delta}]^{\sum_{i=1}^{d}\frac{n_i^2l_s^2}{R_i^2}})\nonumber \\
\label{condthree}
\eea 
In figure 1, we plot (dashed line) the radius dependence of the
contact term due to the exchange of 33 string states (first line of
eq. (\ref{condthree})) in the case where only one of the two radii
appearing in the amplitude is kept finite.  In the infinite radius
limit, we obtain: 
\bea 
\A^{contact}&=& - g_s\, l_s^2\,  \left[   {\bar
u}^{(1)}\gamma_{\mu} u^{(2)}  {\bar u}^{(4)}\gamma^{\mu} u^{(3)}\right
] \int_0^1 \frac {dx}{x} (  \frac {1}{ F(x) } -1 )\nonumber \\ && -
g_s\, l_s^2\left[{\bar u}^{(1)}   u^{(4)} {\bar u}^{(2)} u^{(3)} \right ]
\int_0^1 \frac {dx} {1-x} (\frac {1}{F (x)} - \frac {\pi} {\ln {\frac
{\delta}{1- x}}})
\label{threesets}
\eea 
Performing a change of variables, the above integrands can be
expressed as a function of $\tau$ and the integrals can be evaluated
numerically with the result: 
\bea  
\A^{contact} &\simeq&  {g_s}\,
l_s^2\, \delta_{j_1,{\bar j_2}}  \delta_{j_3,{\bar j_4}} \delta_{{\bar
I_1}, I_4} \delta_{I_2,{\bar I_3}}   \, \{\, 0.12 \, {\bar u}^{(1)}
u^{(4)}  {\bar u}^{(2)} u^{(3)}  + 0.33\, {\bar u}^{(1)} \gamma_{\mu}
u^{(2)}  {\bar u}^{(4)} \gamma^{\mu} u^{(3)}  \,    \} \nonumber \\ &+&
   {g_s}\, l_s^2\, \delta_{I_1,{\bar I_2}} \delta_{I_3,{\bar I_4}}
\delta_{{\bar j_1}, j_4} \delta_{j_2,{\bar j_3}}   \, \, \{\,   0.33\,
{\bar u}^{(1)} \gamma_{\mu} u^{(4)}  {\bar u}^{(2)}  \gamma^{\mu}
u^{(3)}  + 0.12\, {\bar u}^{(1)}  u^{(2)}  {\bar u}^{(4)} u^{(3)}  \,
\}\nonumber \\
\label{approx3}
\eea
where the leading contributions (coefficient $0.33$) are due to the 
exchange of massive oscillators of 33 open strings. On the other hand, in the
limit $R/ l_s \to 0$, one finds the divergent behavior $-\pi^2l_s^2/3R^2$, as in the case
of two sets of branes.

\section{Summary of the effective field theory results}

In this section we would like to summarize the results for the effective 
contact interactions obtained in the previous sections.

In the case where the four-fermion scattering involves two or four states 
arising from DD open strings with both ends on the same set of branes, we have found that the final amplitude can be written as:
\bea
\A (s,t) = \A_{point} (s,t) \cdot \F(s,t) \ , 
\eea
where $\A_{point}$ is the result of the (two derivative) low energy effective action,
while
\bea
   \F(s,t) = {\G(1-l_s^2 s) \G(1-l_s^2 t) \over \G(1-l_s^2s-l_s^2t)} 
= 1 - {\pi^2\over 6}  {st\over M_S^4} + \cdots  
\label{summary1}
\eea
represents the string form factor correction. 
This implies that the leading contact term between the four fermions is 
of dimension-eight. The case of four DD strings can be understood from the 
fact that the tree-level interactions are obtained by a truncation of the 
$N=4$ 
supersymmetric theory arising on D3 branes, upon compactification of 
the ten-dimensional theory. Supersymmetry relates the 
DD fermions to gauge fields and there is no dimension-six operator because of the absence
of $F^3$ terms with $F$ the gauge field strength.

The cases involving four ND strings, describing states localized on brane intersections,
have a richer structure. We have found that the effective field theory contains
dimension-six four-fermion contact terms which get contributions from three different
sources:

\begin{itemize} 

\item {\it Exchange of KK excitations:} Consider for simplicity the case
of two sets of branes, one corresponding to D3 branes and the other to D7 branes. Then,
the exchange of KK excitations of 77 (bulk) states leads to dimension-six operators of the
form:
\bea
- \left[   {\bar \psi}^{(1)} \gamma_{(6)M} \psi^{(2)} 
{\bar \psi}^{(4)} \gamma^{(6)M} \psi^{(3)} \right ] \, \,  
\frac {g_s}{ l_s^{-4}\, \prod_{i=1}^{4} R_i}
 \sum_{m_i \in {\bf Z}-\{ 0\}} 
\frac {e^{i  2 \pi \sum_{i=1}^{4} m_i a_i^{I_1I_3}} \, \, 
\delta^{  - \sum_{i=1}^{4} \frac{m_i^2 l_s^2}{ R_i^2}}}
{ \sum_{i=1}^{4} \frac{m_i^2 }{ R_i^2}}\, , \nonumber \\
\label{kkcon}
\eea
where $\delta=2^4$ and 
$\gamma_{(6)}^{M}$ stand for the $\gamma$-matrices in the six-dimensional spacetime
defined by the directions that are parallel or transverse simultaneously to the D3  and
D7 branes. Reduction to four dimensions leads to 
both gauge and Yukawa couplings. This expression is identical 
to the one obtained in heterotic orbifolds with large extra dimensions \cite{AB}.

The main features of the above result are: (i) the suppresion of the gauge coupling by the
internal volume felt by the bulk states propagating on the D7 branes; 
(ii) the exponential suppression of the coupling of KK excitations (with masses
$m_{KK}$) to the massless localized fermions by the factor
$\delta^{- m_{KK}^2/2M_s^2}$. This factor can be interpreted as a finite width for a
D3 brane, $\sigma=\sqrt{\ln {\delta}}\,  l_s \sim 1.66 \, l_s$ [see eq. (\ref{brwidth})];
(iii) the appearance of phases 
$e^{2i\pi \sum_{i=1}^{4} m_i a_i^{I_1I_3}}$ when  the localized fermions are
separated by the distance $2 \pi a_i^{I_1I_3} R_i$ along the $i$-th direction
\cite{AB}.

\item {\it Exchange of winding modes of the 33 states:} These lead to an operator 
of the form: 
\bea
-   \left[ {\bar \psi}^{(1)} \gamma^{(6)}_{M} \psi^{(4)} 
{\bar \psi}^{(2)} \gamma^{(6)M} \psi^{(3)}   \right ] \, \, g_s \,
 \left[ \sum_{n_i \in {\bf Z}} \frac 
{\delta^{-   \sum_{i=1}^{4} (n_i +a_i^{I_1I_3})^2\,  R_i^2 \, l_s^{-2} }}
{   \sum_{i=1}^{4} (n_i +a_i^{I_1I_3})^2\,  R_i^2 \, l_s^{-4}} \right]\, ,
\label{oper}
\eea 
where in the case of vanishing Wilson lines $a_i^{I_1I_3} = 0$,
the pole due to the exchange of massless states should be substracted
to obtain the contact term. The operator of (\ref{oper}) is exponentially
suppressed in the large radius limit, due to the exchange of massive open
strings streched between the D3 branes.

\item {\it Exchange of massive oscillators:} Their contribution can be simplified by taking
the large (internal) volume limit in the absence of Wilson lines. It is also useful to
separate the contributions associated with the exchanges of 33 and 77 string states.
In the case where all four fermions arise from open strings ending on the same sets of D3 
and D7 branes, we find:
\bea
 \A^{contact}_{33} = - g_s\, l_s^2\,  \int_0^1 \frac {dx}{x} ( 
\frac {1}{ [F (x)]^2 } -1 ) \nonumber \\
\{\delta_{j_1,{\bar j_2}} \delta_{j_3,{\bar j_4}}
\delta_{{\bar I_1}, J_4} \delta_{I_2,{\bar I_3}} 
  &&
 \left[   {\bar \psi}^{(1)} \gamma^{(6)}_{M} \psi^{(2)}  {\bar \psi}^{(4)}
\gamma^{(6)M} \psi^{(3)} \right ]   
 \nonumber \\
- \delta_{I_1,{\bar I_2}} \delta_{I_3,{\bar I_4}}
\delta_{{\bar j_1}, j_4} \delta_{j_2,{\bar j_3}}    &&
 \left[ {\bar \psi}^{(1)} \gamma^{(6)}_{M} \psi^{(4)} 
{\bar \psi}^{(2)} \gamma^{(6)M} \psi^{(3)}   \right ] \} \, \, 
\eea
where $I_a$ and $j_a$ are indices labelling, respectively, the D3  and D7 
branes. A numerical estimate gives:
 \bea 
 \A^{contact}_{33} \simeq 0.59\, {g_s}\, l_s^2\, \{
&& \delta_{j_1,{\bar j_2}} 
\delta_{j_3,{\bar j_4}}
\delta_{{\bar I_1}, I_4} \delta_{I_2,{\bar I_3}}  
\, {\bar \psi}^{(1)} \gamma^{(6)}_{M} \psi^{(2)}  {\bar \psi}^{(4)}
\gamma^{(6)M} \psi^{(3)} \nonumber \\  + && \delta_{I_1,{\bar I_2}} \delta_{I_3,{\bar I_4}}
\delta_{{\bar j_1}, j_4} \delta_{j_2,{\bar j_3}}  
\, \,  {\bar \psi}^{(1)} \gamma^{(6)}_{M} \psi^{(4)}  {\bar \psi}^{(2)} 
\gamma^{(6)M} \psi^{(3)} \}
\label{cont33}
\eea
In the channel corresponding to exchange of 77 states we obtain:
\bea 
 \A^{contact}_{77} = && -g_s \, l_s^2\, \int_0^1 \frac {dx}
{1-x} (\frac {1}{ [F (x)]^2 } - [\frac {\pi} {\ln {\frac {\delta}{1- x}}}]^2)
\nonumber \\ \{ && \delta_{j_1,{\bar j_2}} 
\delta_{j_3,{\bar j_4}}
\delta_{{\bar I_1}, I_4} \delta_{I_2,{\bar I_3}}
{\left[ {\bar \psi}^{(1)} \gamma^{(6)}_{M} \psi^{(4)}  {\bar \psi}^{(2)} \gamma^{(6)M}
\psi^{(3)}
\right ]}
 \nonumber \\ 
+ && \delta_{I_1,{\bar I_2}} \delta_{I_3,{\bar I_4}}
\delta_{{\bar j_1}, j_4} \delta_{j_2,{\bar j_3}}  
    {\left[   
{\bar \psi}^{(1)} \gamma^{(6)}_{M} \psi^{(2)}  {\bar \psi}^{(4)}
\gamma^{(6)M} \psi^{(3)} \right ]} \} 
\label{sum23}
\eea
which can be approximated by:
\bea 
 \A^{contact}_{77} \simeq  0.20 g_s \, l_s^2\, \{ && \delta_{j_1,{\bar j_2}} 
\delta_{j_3,{\bar j_4}}
\delta_{{\bar I_1}, I_4} \delta_{I_2,{\bar I_3}}
 {  \left[ {\bar \psi}^{(1)} \gamma^{(6)}_{M} \psi^{(4)}  {\bar \psi}^{(2)} \gamma^{(6)M} \psi^{(3)}
\right ]}
 \nonumber \\ 
+ && \delta_{I_1,{\bar I_2}} \delta_{I_3,{\bar I_4}}
\delta_{{\bar j_1}, j_4} \delta_{j_2,{\bar j_3}}  
    {\left[   
{\bar \psi}^{(1)} \gamma^{(6)}_{M} \psi^{(2)}  {\bar u}^{(4)}
\gamma^{(6)M} \psi^{(3)} \right ]} \} 
\label{appsum}
\eea

Another possibility arises when there are two different sets of
D7 branes, say D7 and D7', so that one pair of fermions corresponds to
37-strings stretched between the D3  and D7 branes while the other
pair corresponds to 37' strings. In this case, the contact
interactions due to 33-exchanges are: 
\bea 
\A^{contact}_{33} = - g_s
\, l_s^2\, \int_0^1 \frac {dx}{x} (  \frac {1}{ F (x) } -1 ) \nonumber
\\ \{\delta_{j_1,{\bar j_2}} \delta_{j_3,{\bar j_4}} \delta_{{\bar
I_1}, J_4} \delta_{I_2,{\bar I_3}}  && \left[   {\bar \psi}^{(1)}
\gamma_\mu \psi^{(2)}  {\bar \psi}^{(4)} \gamma^\mu \psi^{(3)} \right
]    \nonumber \\ - \delta_{I_1,{\bar I_2}} \delta_{I_3,{\bar I_4}}
\delta_{{\bar j_1}, j_4} \delta_{j_2,{\bar j_3}}    && \left[ {\bar
\psi}^{(1)} \gamma_\mu \psi^{(4)}  {\bar \psi}^{(2)} \gamma^\mu
\psi^{(3)}   \right ] \} \, \,  
\eea 
which can be numerically evaluated: 
\bea  
\A^{contact}_{33} \simeq  0.33 {g_s}\, l_s^2\,
\{ && \delta_{j_1,{\bar j_2}}  \delta_{j_3,{\bar j_4}} \delta_{{\bar
I_1}, I_4} \delta_{I_2,{\bar I_3}}   \, {\bar \psi}^{(1)} \gamma_{\mu}
\psi^{(2)}  {\bar \psi}^{(4)} \gamma^{\mu} \psi^{(3)} \nonumber \\  +
&& \delta_{I_1,{\bar I_2}} \delta_{I_3,{\bar I_4}} \delta_{{\bar j_1},
j_4} \delta_{j_2,{\bar j_3}}   \, \,  {\bar \psi}^{(1)} \gamma^{\mu}
\psi^{(4)}  {\bar \psi}^{(2)}  \gamma_{\mu} \psi^{(3)} \}
\label{cont333}
\eea
Similarly, the exchange of 77' states gives:
\bea 
 \A^{contact}_{77'} =  && -g_s\, l_s^2\, \int_0^1 \frac {dx}
{1-x} (\frac {1}{F (x) } - \frac {\pi} {\ln {\frac {\delta}{1- x}}})
\nonumber \\ \{ && \delta_{j_1,{\bar j_2}} 
\delta_{j_3,{\bar j_4}}
\delta_{{\bar I_1}, I_4} \delta_{I_2,{\bar I_3}}
 {  \left[ {\bar \psi}^{(1)}  \psi^{(4)}  {\bar \psi}^{(2)} \psi^{(3)}
\right ]}
 \nonumber \\ 
+ && \delta_{I_1,{\bar I_2}} \delta_{I_3,{\bar I_4}}
\delta_{{\bar j_1}, j_4} \delta_{j_2,{\bar j_3}}  
    {\left[   {\bar \psi}^{(1)} \psi^{(2)}  {\bar \psi}^{(4)} \psi^{(3)} \right ]} \} 
\label{sum90}
\eea
which can be approximated by:
\bea 
 \A^{contact}_{77'} \simeq  0.12\, g_s \, l_s^2\, \{ && \delta_{j_1,{\bar j_2}} 
\delta_{j_3,{\bar j_4}}
\delta_{{\bar I_1}, I_4} \delta_{I_2,{\bar I_3}}
 {  \left[ {\bar \psi}^{(1)} \psi^{(4)}  {\bar \psi}^{(2)} \psi^{(3)}
\right ]}
 \nonumber \\ 
+ && \delta_{I_1,{\bar I_2}} \delta_{I_3,{\bar I_4}}
\delta_{{\bar j_1}, j_4} \delta_{j_2,{\bar j_3}}  
    {\left[   
{\bar \psi}^{(1)}  \psi^{(2)}  {\bar \psi}^{(4)} \psi^{(3)} \right ]} \}\, . 
\label{xyz}
\eea

\end{itemize}

It is important to notice that the contractions of Chan-Paton indices
in the above formulae are associated with exchanges of $U(N)$ states
and not of $SU(N)$. In a realistic model some of the $U(1)$ factors
are anomalous. The anomalies are canceled by a generalized
Green-Schwarz mechanism which provides the corresponding $U(1)$ gauge
bosons with masses of the order of the string scale~\cite{gs}. While these
$U(1)$'s do not appear in the low energy degrees of freedom, they
contribute in general to the effective four-fermion operators.

In the generic case with finite radii, one can substract the effects
of exchanges of KK excitations that are easily computed in the field
theory limit, to find an effective contact interaction which contains
the combined effects of massive string oscillators and winding
modes. The result is ploted in figure 1  as a function of one
compactification radius. We observe that the size of the operator goes
quickly to its asymptotic value of infinite volume, already for a
radius of order $R\sim 1.4\, l_s$. On the other hand, in the limit
$R/l_s \to 0$, one recovers the asymptotic value due to the sum of the
KK exchanges of masses $mR/l_s^2$. This is given in eq. (\ref{kkcon})
by replacing $R\to l_s^2/R$ (T-duality).

\section{Experimental bounds from  contact terms }

Above, we have shown that the exchange of massive open string states
leads to dimension-six effective operators among four fermions
localized on brane intersections. These operators are generically
parametrized as \cite{fourferm}:
\bea
\L_{eff} = \frac{4 \pi}{(1+\varepsilon )\Lambda^2} \sum_{a,b=L,R}\eta_{ab} 
{\bar \psi_a} \gamma^\mu \psi_a {\bar {\psi'_b}} \gamma_\mu \psi'_b
\label{Lambda}
\eea 
with $\varepsilon=1$ (0) for $\psi=\psi'$ ($\psi \neq \psi'$),
where $\psi_a$ and $\psi'_b$  are left ($L$) or right ($R$) handed
spinors. $\Lambda$ is the scale of contact interactions  and
$\eta_{ab}$ parametrize the relative strengths of various helicity
combinations. In D-brane models there are in general three types of
contact terms due to the exchanges of either massive KK excitations,
or massive winding modes or string oscillator states.

In order to have a clear identification of longitudinal and transverse
directions with respect to the D-branes, we choose all internal radii
to be larger than the string length.  For concretness, we first
consider the simplest scenario where the standard model degrees of
freedom arise from D3 branes with matter fields coming from 37 open
strings having only one end on the D3 branes.  Then, the bulk 77
states have new (exotic) quantum numbers and interact with the 37
matter fields through volume suppressed couplings. Since these
interactions are model dependent we will not consider them  in
our subsequent analysis. The remaining contact terms originate from
the exchange of either massive winding modes of the 33 states or
string oscillator modes. The winding
contributions are exponentially suppressed in the large (internal)
volume limit, and we are left with the contribution of the massive
string states. The result is given in eqs. (\ref{cont33}) and (\ref{appsum})
for the case of two sets of branes and in eqs. (\ref{cont333}) and (\ref{xyz})
for the case of three sets.

For $\psi\neq\psi'$ one can choose the quantum numbers
(i.e. Chan-Paton indices) of the 37 strings such that the operator
(\ref{Lambda}) receives contributions only from the exchange of the
massive 33 open string states. Indeed, by choosing the indices so that
only the first contraction of eqs. (\ref{cont33}) and (\ref{appsum}) survives,
and identifying $u^{(1)}$, $u^{(2)}$ with $\psi$ and $u^{(3)}$, $u^{(4)}$
with $\psi'$, one obtains: 
\bea  
\L_{eff}^{contact} &\simeq&
\frac{g_s}{M_s^2}\, \, \left\{ 0.59 \sum_{a,b=L,R}\eta_{ab}  {\bar
\psi_a} \gamma^\mu \psi_a {\bar {\psi'_b}} \gamma_\mu \psi'_b + 0.20
\sum_{a,b=L,R}\eta_{ab}  {\bar \psi_a} \gamma^\mu \psi'_a {\bar
{\psi_b}} \gamma_\mu \psi'_b \right\} \nonumber\\
\label{psi}
\eea 
where the two terms correspond to the exchanges of 33 and 77 open
string states, respectively. Thus, the first operator which coincides
with (\ref{Lambda}) receives contributions only from the exchange of
massive 33 open string oscillators. Exchanges of 77 states contribute
 to the second operator which induces new (flavor changing)
interactions that are model dependent and must be suppressed in a
realistic model. In fact, it is easy to see that there is a choice of
spinor helicities ($\psi_L, \psi_L, \psi'_R, \psi'_R$) which
eliminates the second operator completely. The same analysis holds for
the case where the external fermions arise from three different sets
of branes with eq. (\ref{psi}) replaced by the first line of
eqs. (\ref{cont333}) and (\ref{xyz}).

It follows that one can identify the parameters in eq. (\ref{Lambda}) as: 
\bea 
\eta_{LL}= \eta_{RR}&=&\eta_{LR}=\eta_{RL}= 1\\
\Lambda &\simeq& \sqrt{\frac{4\pi}{{0.59} g_s}} M_s
\label{fres1}
\eea
in the case of two sets of banes, while for 3 sets 0.59 is replaced by
0.33.  The signs and relative ratios of the different terms in
(\ref{Lambda}) correspond to what is usually refered to as
$\Lambda_{VV}^{+}$. The present bounds from LEP \cite{LEP} are of  the order of
$\Lambda_{VV}^{+} \simgt 16$ TeV which for $g_s=g_{YM}^2 \sim 1/{2}$,
with $g_{YM}$ the gauge coupling, leads to  $M_s \simgt 2.5$ TeV while
it becomes $M_s \simgt 1.1$ TeV for the extreme choice  $g_s/4\pi =
1/128$.  The later bound is to be compared with  the limit obtained
from dimension-eight operators that arise in the case of fermions from
DD strings  $M_s \simgt 0.63$ TeV \cite{Peskin,lim}. In the case of
three sets of branes, the limits become $M_s \simgt 1.8$ and $0.81$ TeV 
for the two respective values of $g_s$.

A stronger bound can be obtained from the analysis of high precision low
energy data in the presence of effective four-fermion operators that
modify the $\mu$-decay amplitude. They are obtained from the first term
of eq.~(\ref{psi}) with $\psi$ and $\psi'$ the muon and electron
doublets, respectively, and with an overall negative sign. Using the
results of ref.~\cite{strumia}, we obtain $M_s \simgt 3.1$ TeV for
$g_s\simeq 1/{2}$, or $M_s \simgt 2.8$ TeV for $g_s=g_2^2\simeq 0.425$
with $g_2$ the $SU(2)$ coupling at the weak scale. Note that in this case
the ambiguity on the value of the coupling is not important. In the case
of three sets of branes, the limits become $M_s \simgt 2.2$ TeV.

In the case $\psi = \psi'$ as for Bhabha scattering in $e^+ e^-$
there is an  additional contribution to the effective operator coming
from the operators that are  associated with the exchange of 77
states. In the case of two sets of branes, this leads to: 
\bea
0.75\, \eta_{LL}=0.75\, &\eta_{RR}&\simeq\eta_{LR}=\eta_{RL}= 1\\
\Lambda &\simeq& \sqrt{\frac{4\pi}{{0.59} g_s}} M_s
\label{fres2}
\eea
where we see that the dominant effects are still due to the 33 channels.

Finally, for comparison, we would like to consider the case where
Standard Model gauge interactions appear in a higher dimensional
D-brane and feel the presence  of additional longitudinal
dimensions. This can be obtained either by  identifying observable
gauge bosons with D7 brane states or with those of D4 branes obtained from the
previous D3 branes by taking one of the transverse 
directions in the world-volume
of D7 branes smaller than the string length and performing a
T-duality. The resulting  contact interaction is dominated by the
effects due to exchange of KK excitations. These  lead to an operator
of the form  (\ref{kkcon}). In the case of one transverse
dimension, the sum over KK states gives \cite{AB}: 
\bea 
\L_{eff}^{KK}
\simeq -\frac{ \pi^2}{3(1+\varepsilon )} R^2 g_s
\sum_{a,b=L,R}\eta_{ab}  {\bar \psi_a} \gamma^\mu \psi_a {\bar
{\psi'_b}} \gamma_\mu \psi'_b
\label{Lambda2}
\eea Experimental constraints on such operators translate into lower
bounds on the scale of compactification. For instance exchanges
leading to vector interactions would lead to: 
\bea  
\eta_{LL}=
\eta_{RR}&=&\eta_{LR}=\eta_{RL}= - 1\\ \Lambda &\simeq&
\sqrt{\frac{12}{\pi g_s}} \frac{1}{R}
\label{fres3}
\eea 
Using LEP bounds \cite{LEP} $\Lambda_{VV}^{-} \simgt 14$ TeV and
$g_s=g_{YM}^2 \sim 1/{2}$, with $g_{YM}$ the gauge coupling, gives
$R^{-1} \simgt 5.0$ TeV. This bound  becomes $R^{-1} \simgt 2.2$ TeV
for the choice  $g_s/4\pi = 1/128$ corresponding to contact terms
dominated for instance by the exchange of KK excitations of photons.
Low energy precision electroweak data lead instead to 
$R^{-1} \simgt 3.5$ TeV \cite{KK2}.

\section*{Acknowledgements}
We thank Alessandro Strumia for useful comments on extracting
experimental bounds using LEP1 data for charged currents.
This work was supported in part by the European Commission under TMR
contract ERBFMRX-CT96-0090 and RTN contract HPRN-CT-2000-00148, and in
part by the INTAS contract 99-0590. The work of K.B. was also partially
supported by NSERC of Canada and  Fonds FCAR du Qu\'ebec.

\appendix
\section{ Four-scalar scattering amplitudes from  ND open strings}

 The massless scalars arising from open strings with  ND boundary
conditions transform in spinorial representations of  the $SO(4)$
internal symmetry corresponding to the four twisted directions.  The
corresponding vertex operators are given (in the -1 ghost picture) by:
\be 
{\cal V}^{(a)}_{ND} (x_a,k_a) =  \sqrt { 2 g_s}\, l_s\,   e^{i
k_i\cdot X} \, \lambda_{j_aI_a} \, e^{-\phi}\,  \eta^{(a)}_\alpha
S^\alpha_{(4)} \,  \prod_{j=1}^4 \sigma^j_{\pm} \,  (x_a)\, ,  
\ee 
where
the momenta $k_a$ have components only in the directions  longitudinal
to the  D3 brane and satisfy  $k_a^2 =0$. The  $S^\alpha_{(4)}$ and
$\eta^{(a)}_\alpha $ represent the spin field and the corresponding
polarization in the internal $SO(4)$, respectively. The computation of
the ordered amplitude involving four scalars from ND  open
strings is straightforward and  leads to: 
\bea \A (1_{j_1 I_1},2_{j_2
I_2},3_{j_3 I_3},4_{j_4 I_4})&=&    -  2 u { g_s} \, l_s^2\, \int_0^1 dx \, \,
x^{-1 -s\, l_s^2}\, \, \,   (1-x)^{-1 -t\, l_s^2} \, \, \,  \frac {1}{
[F (x)]^2 } \nonumber \\  & \times&  \left[   {\bar \eta}^{(1)}
\Gamma_{a} \eta^{(2)}  {\bar \eta}^{(4)} \Gamma^{a} \eta^{(3)} (1-x) +
{\bar \eta}^{(1)} \Gamma_{a}  \eta^{(4)} {\bar \eta}^{(2)} \Gamma^{a}
\eta^{(3)}  x \right ]  \nonumber \\  & \times& \{  \frac
{\delta_{I_1,{\bar I_2}} \delta_{I_3,{\bar I_4}} \delta_{{\bar
j_1}, j_4} \delta_{j_2,{\bar j_3}}}{l_s^{-4} \, \prod_{i=1}^{4} R_i\, }
\sum_{m_i} {e^{i  2 \pi \sum_i m_i a_i^{I_1I_3}} \, \, e^{ - {\pi}
{\tau}\, \sum_i \frac {  m_i ^2 \, l_s^2 }{ R_i^2}   }} \nonumber \\
&& +   \delta_{j_1,{\bar j_2}}  \delta_{j_3,{\bar j_4}} \delta_{{\bar I_1},
I_4} \delta_{I_2,{\bar I_3}} \sum_{n_i}  e^{-  {\pi \tau}
\sum_i (n_i +a_i^{I_1I_3})^2\,  R_i^2 \, l_s^{-2} } \, \, \, \, \}
\label{mastereq2}
\eea
where $n_i$ and $m_i$ are integers. The total amplitude is obtained by 
summing up all orderings.

The  low-energy limit $|s \, l_s^2| \ll 1$ and $|t\,  l_s^2| \ll 1$  
of the amplitude in (\ref{mastereq2}) can be computed and is decomposed in
two terms $\A^{QFT}_{1}$ and $\A^{QFT}_{2}$:
\bea 
 \A^{QFT}_{1} &=& - \, g_s\,
\delta_{j_1,{\bar j_2}} 
\delta_{j_3,{\bar j_4}}
\delta_{{\bar I_1}, I_4} \delta_{I_2,{\bar I_3}} 
\nonumber \\
&\times&   \{  (s-u) \frac {  \left[ {\bar\eta
}^{(1)} \Gamma_{a} \eta^{(4)}  {\bar \eta}^{(2)} \Gamma^{a} \eta^{(3)}
\right ]} { l_s^{-4} \prod_{i=1}^{4} R_i} 
\, \left[ \sum_{m_i } \frac
{e^{i  2 \pi \sum_i m_i a_i^{I_1I_3}} \, \, \delta^{  - \sum_i \frac{m_i^2
l_s^2}{ R_i^2}}}{t - \sum_i \frac{m_i^2}{ R_i^2}} \right] \nonumber \\ 
&+& (t-u)
 \left[   {\bar \eta}^{(1)} \Gamma_{a} \eta^{(2)}  {\bar \eta}^{(4)}
\Gamma^{a} \eta^{(3)} \right ] 
\left[\sum_{|n_i +a_i^{I_1I_3}|\,  R_i \, < \, l_s} \frac 
{\delta^{-   \sum_i (n_i +a_i^{I_1I_3})^2\,  R_i^2 \, l_s^{-2} }}
{s  -   \sum_i (n_i +a_i^{I_1I_3})^2\,  R_i^2 \, l_s^{-4}} \right] \, \,  \, \, 
\} \nonumber \\
&-&    g_s\,
 \delta_{I_1,{\bar I_2}} \delta_{I_3,{\bar I_4}}
\delta_{{\bar j_1}, j_4} \delta_{j_2,{\bar j_3}} \nonumber \\
&\times&   \{  (t-u) \frac{\left[   {\bar \eta}^{(1)} \Gamma_{a} \eta^{(2)}  {\bar \eta}^{(4)}
\Gamma^{a} \eta^{(3)} \right ]} { l_s^{-4}\,
\prod_{i=1}^{4} R_i}  \left[ \sum_{m_i} \frac {e^{i  2 \pi \sum_i m_i a_i^{I_1I_3}} \, \, \delta^{  - \sum_i \frac{m_i^2 l_s^2}{ R_i^2}}}{s - \sum_i \frac{m_i^2}{ R_i^2}} \right] \label{usual} \\&+& (s-u) \left[ {\bar
\eta}^{(1)} \Gamma_{a} \eta^{(4)}  {\bar \eta}^{(2)} \Gamma^{a} \eta^{(3)}
\right ]
 \left[ \sum_{|n_i +a_i^{I_1I_3}|\,  R_i \, < \, l_s } \frac 
{\delta^{-   \sum_i (n_i +a_i^{I_1I_3})^2\,  R_i^2 \, l_s^{-2} }}
{t  -   \sum_i (n_i +a_i^{I_1I_3})^2\,  R_i^2 \, l_s^{-4}} \right]
\, \,  \}\nonumber 
\eea
and
\bea 
 \A^{QFT}_{2} &=&  -  g_s \, \, 
\delta_{j_1,{\bar j_2}} 
\delta_{j_3,{\bar j_4}}
\delta_{{\bar I_1}, I_4} \delta_{I_2,{\bar I_3}}  \nonumber \\
&\times&   \{   \frac {  \left[ {\bar
\eta}^{(1)} \Gamma_{a} \eta^{(4)}  {\bar \eta}^{(2)} \Gamma^{a} \eta^{(3)}
\right ]} { l_s^{-4} \prod_{i=1}^{4} R_i} 
\, \left[ \sum_{m_i } 
{e^{i  2 \pi \sum_i m_i a_i^{I_1I_3}} \, \, \delta^{  - \sum_i \frac{m_i^2
l_s^2}{ R_i^2}}} \frac {t}{t - \sum_i \frac{m_i^2}{ R_i^2}} \right] \nonumber \\ 
&+& 
 \left[   {\bar \eta}^{(1)} \Gamma_{a} \eta^{(2)}  {\bar \eta}^{(4)}
\Gamma^{a} \eta^{(3)} \right ] 
\left[\sum_{|n_i +a_i^{I_1I_3}|\,  R_i \, < \, l_s}  
{\delta^{-   \sum_i (n_i +a_i^{I_1I_3})^2\,  R_i^2 l_s^{-2} }}
\frac  {s  }
{s  -   \frac{ \sum_i (n_i +a_i^{I_1I_3})^2\,  R_i^2}{l_s^4} } \right]  
\} \nonumber \\
&-& g_s \, \, 
 \delta_{I_1,{\bar I_2}} \delta_{I_3,{\bar I_4}}
\delta_{{\bar j_1}, j_4} \delta_{j_2,{\bar j_3}}  
   \label{new} \\
&\times&   \{   \frac{\left[   {\bar \eta}^{(1)} \Gamma_{a} \eta^{(2)}  {\bar \eta}^{(4)}
\Gamma^{a} \eta^{(3)} \right ]} { l_s^{-4}\,
\prod_{i=1}^{4} R_i} \left[ \sum_{m_i } 
{e^{i  2 \pi \sum_i m_i a_i^{I_1I_3}} \, \, \delta^{  - \sum_i \frac{m_i^2
l_s^2}{ R_i^2}}} \frac {s}{s - \sum_i \frac{m_i^2}{ R_i^2}} \right]
\nonumber \\&+&  \left[ {\bar
\eta}^{(1)} \Gamma_{a} \eta^{(4)}  {\bar \eta}^{(2)} \gamma^{a} \eta^{(3)}
\right ]
\left[\sum_{|n_i +a_i^{I_1I_3}|\,  R_i \, < \, l_s}  
{\delta^{-   \sum_i (n_i +a_i^{I_1I_3})^2\,  R_i^2 l_s^{-2} }}
\frac {t  }
{t  -    \frac {\sum_i (n_i +a_i^{I_1I_3})^2\,  R_i^2 }{l_s^{4}}} \right]
  \, \}
\nonumber
\eea
which reproduce the effective field theory contributions from the exchange 
of gauge bosons and their KK excitations. The result in eq. (\ref{usual}) 
 arises when the derivative, present in the interaction 
vertex,  acts on the  external ND scalar states. Taking $a_i^{I}= 0$, 
 eq. (\ref{new}) contains a sum of terms obtained when  acting by the 
derivative on the internal  massive KK excitation of 77 states. In addition,
it contains
a four-scalar contact term that corresponds to $m_i=n_i=0$
\bea 
 \A^{contact} = &-&  {g_s}  \, 
\, \delta_{j_1,{\bar j_2}} 
\delta_{j_3,{\bar j_4}}
\delta_{{\bar I_1}, I_4} \delta_{I_2,{\bar I_3}}  \{   \frac {   {\bar
\eta}^{(1)} \Gamma_{a} \eta^{(4)}  {\bar \eta}^{(2)} \Gamma^{a} \eta^{(3)}}
 { l_s^{-4} \prod_{i=1}^{4} R_i} 
\,  
+  {\bar \eta}^{(1)} \Gamma_{a} \eta^{(2)}  {\bar \eta}^{(4)}
\Gamma^{a} \eta^{(3)} \, \} \nonumber \\
&-&  {g_s}  \, 
 \delta_{I_1,{\bar I_2}} \delta_{I_3,{\bar I_4}}
\delta_{{\bar j_1}, j_4} \delta_{j_2,{\bar j_3}}  
  \,
   \{   \frac{   {\bar \eta}^{(1)} \Gamma_{a} \eta^{(2)}  {\bar \eta}^{(4)}
\Gamma^{a} \eta^{(3)}} { l_s^{-4}\,
\prod_{i=1}^{4} R_i}  +  {\bar
\eta}^{(1)} \Gamma_{a} \eta^{(4)}  {\bar \eta}^{(2)} \gamma^{a} \eta^{(3)}
 \, \}
\nonumber\\
\label{Dterm}
\eea 
which reproduces the usual quartic coupling  from $D$-terms in
supersymmetric theories. Dimension-six operators involving four saclars can be
extracted easily from the above results and they are subleading.



\end{document}